\documentclass{article}

\usepackage{amsmath,amssymb}
\usepackage[a4paper]{geometry}
\usepackage{graphicx}

\def\be{\begin{equation}}
\def\ee{\end{equation}}

\DeclareMathOperator\erf{erf}
\DeclareMathOperator\erfi{erfi}
\DeclareMathOperator\arctanh{arctanh}

\begin{document}

\title{\bf Phenomenological picture of fluctuations
in branching random walks}

\author{A. H. Mueller${}^1$, S. Munier${}^2$%
\footnote{Corresponding author. Email: {\tt Stephane.Munier@polytechnique.edu}}\\
\\
\small\it ${}^{(1)}$Department of Physics, Columbia University, New York, USA\\
\small\it ${}^{(2)}$Centre de physique th\'eorique, \'Ecole Polytechnique, 
CNRS, Palaiseau, France.
}
\date{}

\maketitle

\begin{abstract}
We propose a picture of the fluctuations in branching random walks, 
which leads
to predictions for the distribution of a random variable
that characterizes the position of the bulk of the particles.
We also interpret the $1/\sqrt{t}$ correction to
the average position of the rightmost particle of a
branching random walk for large times $t\gg 1$,
computed by Ebert and Van Saarloos,
as fluctuations on top of the mean-field approximation of this
process with a Brunet-Derrida cutoff
at the tip that simulates discreteness.
Our analytical formulas successfully compare to numerical simulations
of a particular model of branching random walk.
\end{abstract}


\section{Introduction}

The goal of this work is to better understand the distribution
of the particles generated by a branching random walk process
after some large evolution time.\\

Our initial motivation for addressing this problem comes 
from particle physics~\cite{Iancu:2004es} 
(for a review, see~\cite{Munier:2009pc}).
In the context of the scattering of hadrons at large energies,
high-occupation quantum fluctuations 
dominate some of the scattering cross sections currently measured for example at the LHC.
These quantum fluctuations can be thought of as 
being built up, as the hadrons are accelerated, by the successive branchings 
first of their constituant
quarks into quark-gluon pairs, and then of the
gluons into pairs of gluons, with some diffusion in
their momenta. 
The dynamics of these gluons is 
actually exactly the kind of branching diffusion process that we are going to address
in this work.
Therefore, results that do not depend on the detailed properties 
of the particular branching random walk considered
may be transposed to particle physics, and give
quantitative insight into hadronic scattering cross sections.

Of course, the applications of branching random walks
are much wider than particle physics. 
Branching random walks may for example generate Cayley trees
which would represent the configuration space of directed polymers
in random media \cite{DS:1988}.
\\

\begin{figure}
\begin{center}
\begin{tabular}{cc}
\includegraphics[width=.45\textwidth,angle=0]{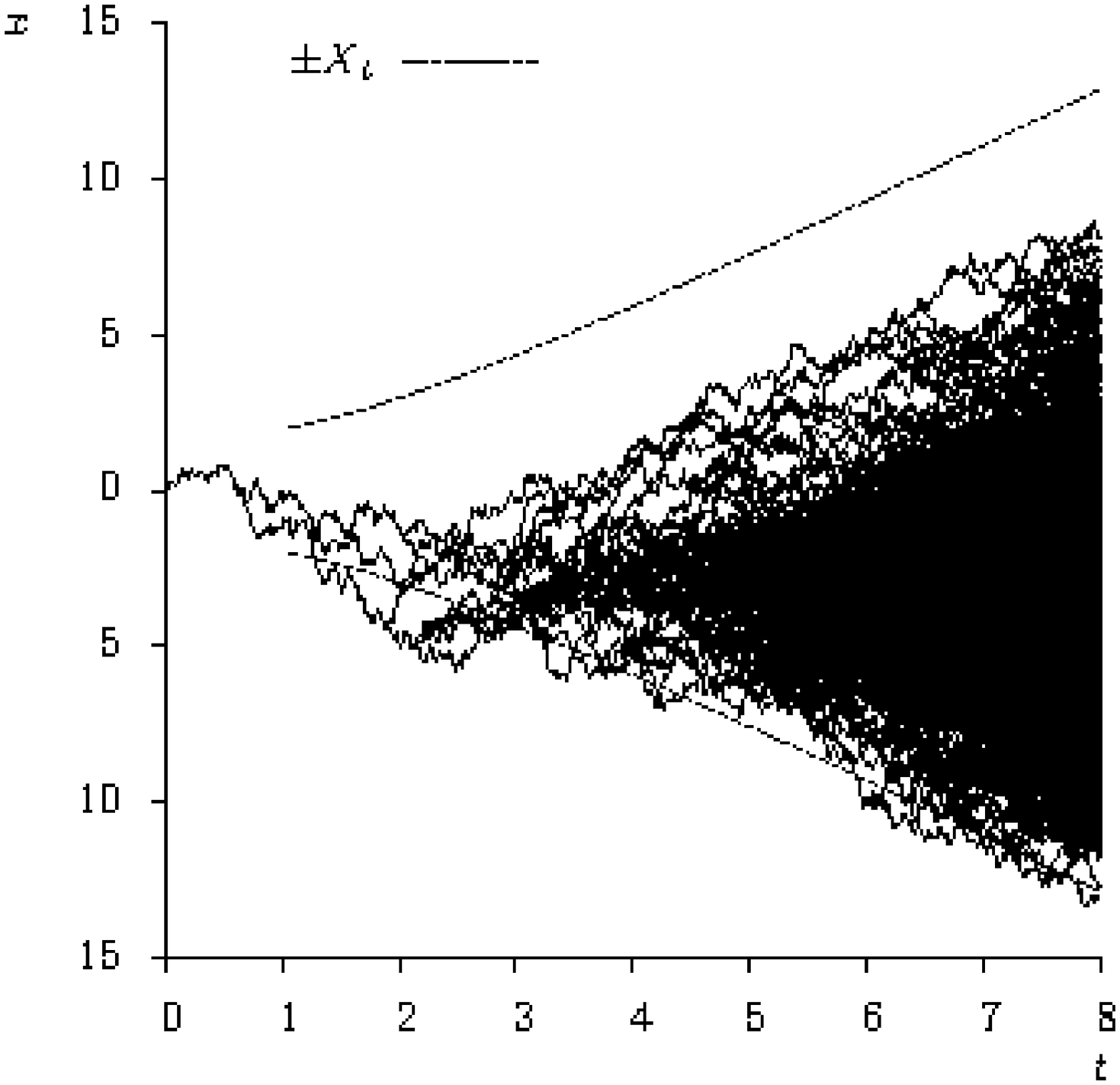} &
\includegraphics[width=.45\textwidth,angle=0]{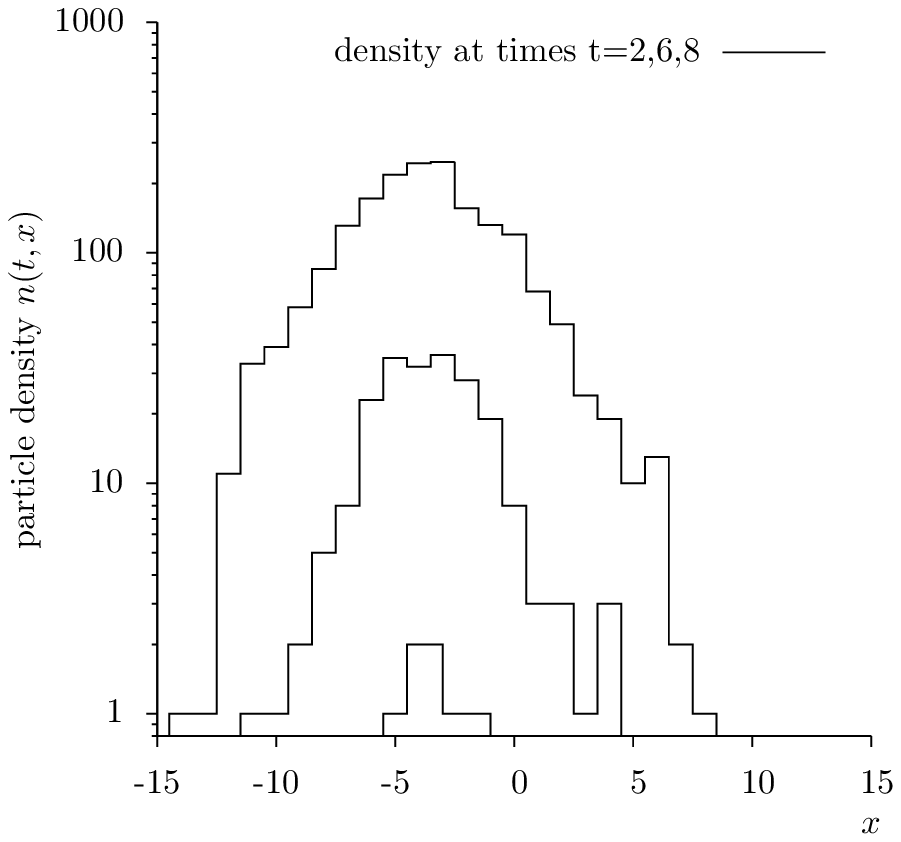}\\
(a) & (b)
\end{tabular}
\end{center}
\caption{\label{fig:plotBRW}{(a)} One realization of the continuous
BRW up to time $t=8$. 
To guide the eye, we also plot the theoretical (truncated)
mean position of the boundaries
of the BRW, namely $\pm\bar X_t=\pm 2t\mp\frac32 \ln t$ (dashed lines).
{(b)} Distribution of the particles at times $t=2,6$ and $8$
for this particular realization in bins of size 1 
($\log_{10}$ scale on the vertical axis).
We see the bulk building up a smoother (more ``deterministic'') distribution
as time elapses, while the low-density tails remain noisy.
Also, for this realization, the distribution is skewed towards negative
values of $x$, due to an accidentally 
large drift in the initial stages, whose memory is kept throughout
the evolution.
}
\end{figure}

Although our discussion will be very
general, for definiteness, we shall
consider a simple model for a branching random walk (BRW)
in continuous time $t$ and one-dimensional space $x$, defined by
two elementary processes: 
Each particle diffuses independently of the others
with diffusion constant 1, and
may split into two particles at rate 1, in such a way that the
mean particle density $\langle n(t,x)\rangle$ obeys the equation
\be
\partial_t \langle n(t,x)\rangle=\partial_x^2 \langle n(t,x)\rangle+\langle n(t,x)\rangle.
\label{eq:meannBRW}
\ee
A particular realization of this BRW is represented in Fig.~\ref{fig:plotBRW}.\\

Several properties of BRW are known.
In particular,
in any given realization of the stochastic process,
for large enough times, the forward part of the distribution of the particles
looks like an exponential $e^{-x}$ 
(scaled by an appropriate time-dependent constant, also
depending on the particular
realization considered)
up to fluctuations effectively concentrated at its low-density tip.
We shall call this exponential part the ``front.''

Then, one can also establish 
rigorously~\cite{McK:1975,B:1983}
that the probability $Q(t,x)$ that all particles
sit at a position smaller than $x$ obeys a nonlinear partial differential
equation which reads
\be
\partial_t Q(t,x)=\partial_x^2 Q(t,x)-Q(t,x)+Q^2(t,x).
\label{eq:FKPP}
\ee
This is a version of
the Fisher-Kolmogorov-Petrovsky-Piscounov (FKPP) equation \cite{F:1937,KPP:1937}.
(For an extensive review, see Ref.~\cite{VS:2003}, and for more
applications of the FKPP equation, see e.g. Ref.~\cite{Majumdar:2002}).
If the BRW starts at time $t=0$ with a single particle located at $x=0$, then the initial
condition is $Q(t=0,x)=\theta(x)$.

With such an initial condition, the solution of the FKPP equation tends to
a so-called ``traveling wave''.
The position of a FKPP traveling wave, which is related to 
the average position of the rightmost particle
in the BRW, is known in the large-time limit:
\begin{equation}
\left[\text{FKPP front position}\right]=2t-\frac{3}{2}\ln t
+\text{const}
+\frac{C_X}{\sqrt{t}}+\cdots\ \ 
\text{with}\ \ 
C_X=-3\sqrt{\pi},
\label{eq:FKPPpos}
\end{equation}
where the last term was found by Ebert and Van Saarloos \cite{EVS:2000}.
(The additive constant depends on the way one defines the position of the front.
It is uninteresting for our purpose.)
Note that the Ebert-Van Saarloos term is a decreasing but {\em positive}
contribution to the front velocity.
Equation~(\ref{eq:FKPPpos}) 
may easily be extended to different branching diffusion models
by appropriately replacing some numerical constants
(see below, Sec.~\ref{sec:numerics}).

More generally, if $N(t)$ is the number of particles at time $t$, and $\{x_i(t)\}$ is the
set of their positions in a given realization, then
\be
G_t(x)\equiv\left\langle \prod_{i=1}^{N(t)} f(x-x_i(t))\right\rangle
\label{eq:G}
\ee
for any given function $f$ satisfies the same FKPP equation as $Q$,
the initial condition 
being the function $f(x)$ itself in the case
of a BRW starting with one single particle at the origin.
If $f$ is a monotonous function of $x$
such that $f(x)\underset{x\rightarrow-\infty}{\longrightarrow}0$
and $f(x)\underset{x\rightarrow+\infty}{\longrightarrow}1$, and if $f$ 
reaches 1 fast
enough, namely $1-f(x)\underset{x\rightarrow+\infty}{\sim}e^{-\gamma x}$
with $\gamma>1$, then the traveling wave solution holds and the  front position
is still given by Eq.~(\ref{eq:FKPPpos}).
Another interesting particular
case is the ``critical case'' 
when $f$ is such that $\gamma=1$ exactly. Then,
\begin{equation}
\left[\text{FKPP front position}\right]_\text{critical}=2t-\frac{1}{2}\ln t
+\text{const}
+{\frac{C_Y}{\sqrt{t}}}+\cdots
\label{eq:FKPPposcrit}
\end{equation}
where $C_Y$ is a constant that we shall determine later on (see Sec.~\ref{sec:det}).\\

There also exists a theorem established by Lalley and Sellke
\cite{LS:1987}
that gives the asymptotic (large time)
shape of the distribution of the position of the rightmost
particle in a frame whose origin is 
at position $\ln Z$, where $Z$ is some random 
variable that depends on the realization and may be thought of
as a characterization of the position of the bulk of the particles in the BRW.
(Its precise definition will be given later on).
More recently \cite{BD:2009,BD:2011,Aidekon:2013}, the distribution of the 
distances between the foremost particles was derived with the help of the solution to the
FKPP equation with some peculiar initial condition.\\

In this paper, we propose a phenomenological
picture of the fluctuations in BRW,
and we derive within this picture some new statistical properties 
of a random variable similar to $\ln Z$.
(Appendix~\ref{sec:Z} also lists some properties of $\ln Z$ itself.)
\\

In Sec.~\ref{sec:model}, we shall introduce our phenomenological picture
for branching random walks. Section~\ref{sec:distrib} is devoted
to deriving the quantitative predictions of this model for a particular
random variable that can characterize the early-time fluctuations of
the branching random walk.
The computation of a few free constant parameters
requires us to solve deterministic equations:
This is explained in Sec.~\ref{sec:det}.
Numerical checks are in order since our analytical results are based
on conjectures: This is done in Sec.~\ref{sec:numerics}.
In light of our phenomenological model, we shall then come back
to the discussion of the Ebert-Van Saarloos result on the $1/\sqrt{t}$
correction to the position
of FKPP fronts (Sec.~\ref{sec:EVS}).
Conclusions are given in Sec.~\ref{sec:conclusion}.


\section{\label{sec:model}
Phenomenological description of branching random walks}

\subsection{Picture}

The picture of the fluctuations in branching 
random walks (BRWs) that we have 
in mind is the following.
There are essentially three types of fluctuations
that may affect the position of the front or of the foremost particle.

\begin{enumerate}
\item
First, there are fluctuations occurring at very early times 
($t\sim 1$), when the system consists in a few particles. 
They have a large (of order 1) and lasting impact 
on the position of the front or of 
the rightmost particle.
The main effect is given by the random waiting time of the first
particle before it splits into two particles, 
during which it diffuses,
but the subsequent
waiting times of
the latter two particles also contribute, etc... 
until the system contains a large enough number of particles
that makes it partly ``deterministic''.
We do not believe that there is a simple way to compute the effect of 
these fluctuations, since there is no large parameter in the problem
which would allow for sensible approximations.

\item
Once the system contains many particles,
which happens say at time $t_0^\prime\gg 1$,
it enters a ``mean-field'' regime:
In a first approximation,
its particle density obeys a deterministic evolution with a moving absorptive
boundary at a position that we shall call
$\bar X_t$ (and symmetrically at $-\bar X_t$), 
set in such a way that the particle density is 1 at
some fixed distance of order 1 to the left or to the right of this boundary,
respectively.
These boundaries simulate the discreteness of the particles.
This is the Brunet-Derrida cutoff \cite{BD:1997}, and it was shown to correctly represent the
leading effect of the noise on the position of the front in the context
of the stochastic FKPP equation.

From now on, we shall 
focus on the right boundary. (The right and the left halves of the
BRW essentially decouple once the system has grown large enough).
The large-time expression of the shape of the particle density 
near the right boundary
reads, in such a model
\begin{equation}
\psi_{\bar X_t}(x,t)=\left[\alpha(x-\bar X_t)+\beta\right] 
\exp\left(\bar X_t-x-\frac{(\bar X_t-x)^2}{4t}\right)
\theta(\bar X_t-x)
\label{eq:shape}
\end{equation}
in the region $1\ll\bar X_t-x\lesssim \sqrt{t}$,
where
\begin{equation}
\bar X_t=2t-\frac32\ln t+\frac{C_{\bar X}}{\sqrt{t}}
\label{eq:Xt}
\end{equation}
is, up to an uninteresting non-universal additive constant, 
the position of the tip of the front,
namely of the right boundary. 
The Heaviside step function $\theta$ enforces the fact that the particle
density is 0 to the right of $\bar X_t$.
$C_{\bar X}$, $\alpha<0$ and $\beta$
are constants
undetermined at this stage.
$\psi$ is essentially a decreasing exponential supplemented with a Gaussian
and a linear prefactors. The $t$-dependence enters explicitly as the width of the
Gaussian, and implicitly through the position $\bar X_t$ of the absorptive
boundary.
(There are corrections to the shape of the front at order $1/\sqrt{t}$,
namely to the function $\psi_{\bar X}$ itself, but it turns out
that we do not need to take them into account in our model, except for
the determination of one overall
numerical constant: We will come back to the
derivation of Eq.~(\ref{eq:shape}) and 
of its corrections in Sec.~\ref{sec:det}.)

The fluctuations 
on top of this essentially deterministic front we have just
described
must take place in the tip region,
where the particle density is low.
We shall assume that a single fluctuation effectively gives the dominant
correction to the deterministic evolution, 
and that the distribution $p(\delta)$ 
of the position $\delta$ of
this fluctuation with respect to the tip of the front
is exponential:
\begin{equation}
p(\delta)=C_1 e^{-\delta}.
\label{eq:proba}
\end{equation}
We have found (see below) that these fluctuations 
bring a contribution of order $1/\sqrt{t}$ to
the average position both of the front and of 
the rightmost particle in the BRW.

\item
Finally, there are tip fluctuations occurring at very late times, 
say between $t-\bar t_0$ and $t$, 
where $\bar t_0$ is of order~1. They are also distributed as $e^{-\delta}$.
They obviously add noise to the position of the tip of the front, 
but they 
do not have an effect on the bulk of the particle distribution
since they do not have time to develop their own front at time $t$.
\end{enumerate}
This picture is parallel to the phenomenological model for 
front fluctuations proposed in Ref.~\cite{Brunet:2005bz}
in the context of the stochastic FKPP problem.

\subsection{\label{sec:var}Variables}

To arrive at quantitative predictions for the behavior of
the BRW, we need to introduce random variables that characterize the realizations.
We shall discuss the following ones:
\begin{itemize}
\item $X_t$,
the position of the 
rightmost particle,
\item $Y_t=\sqrt{t}\sum_i e^{x_i(t)-2t}$, where the sum goes over all
the particles in the system,
\item $Z_t=\sum_i \left[2t-x_i(t)\right]e^{x_i(t)-2t}$.
\end{itemize}
Throughout, we shall denote by $\langle A \rangle$ the statistical
average (over realizations) of a given variable $A$
in the full stochastic model, 
and by $\bar{A}$ the value of this variable
in a mean-field approximation of the same model with a discreteness
cutoff at the tip.
(These notations have already been used in Eq.~(\ref{eq:meannBRW})
and Eqs.~(\ref{eq:shape}),~(\ref{eq:Xt}) respectively.)
Discrete sums over the particles will often
be replaced by integrals wherever the particle density is large enough.\\

\noindent
Let us briefly comment on the random variables we have just introduced.
\begin{itemize}
\item
As already mentioned,
$\langle X_t\rangle$ is related to the solution
of the FKPP equation with the step function
as an initial condition.
\item
The average $\langle \ln Y_t\rangle$ 
tends to a constant at large $t$. In addition, in any given
event, the random variable $\ln Y_t$ itself tends to a constant, which
has some distribution (which we do not know how to compute) which may
be used to characterize
the early-time fluctuations.
Note that an appropriate 
generating function of
the moments of 
\be
\tilde Y_t\equiv \sum_i e^{x_i(t)}=Y_t\times \frac{e^{2t}}{\sqrt{t}}
\label{eq:defYt}
\ee
also obeys the FKPP equation, but with 
the ``critical'' initial condition
discussed in the Introduction.
We will come back to the latter fact in Sec.~\ref{sec:det}.
Also, in the context of directed polymers in random media,
$\tilde Y_t$ is the partition function and 
$\langle\ln \tilde Y_t\rangle$ the
free energy averaged over the disorder~\cite{DS:1988}.
\item
$Z_t$ is the variable used by Lalley and Sellke
in the theorem alluded to in the Introduction.
However, we are not going to focus on this variable in the
body of this paper, since we found it has
many drawbacks for our purpose. First, a practical drawback: 
Although $Z_t$ tends almost
surely to a positive constant when $t\rightarrow+\infty$ \cite{LS:1987},
it takes negative values
at finite times, 
with finite probability; $\ln Z_t$ is then undefined in these particular
realizations.
Second, a theoretical drawback:
It turns out that the finite-time corrections to the moments of
$\ln Z_t$ are very sensitive
to the initial fluctuations, the ones that are not computable analytically.
We shall nevertheless quote a few results on the distribution 
of $\ln Z_t$ in Appendix~B.
\end{itemize}
In some intuitive sense, $\ln Y_t$ and $\ln Z_t$ characterize the
position of the ``front'' of a particular realization of the evolution at time $t$.\\

The variables $\ln Y_t$ and $\ln Z_t$ keep the memory of the initial
fluctuations.
Therefore, we shall not attempt to compute 
the distribution of the fluctuations in $\ln Y$
accumulated over the whole history of the BRW,
but instead the fluctuations of this variable between
two large times $t_0$ and $t$, in order to have a quantity
that is independent of the very early times at which there
is no mean-field regime.


\section{\label{sec:distrib}Statistics of $f\equiv \ln Y_t-\ln Y_{t_0}$
in the phenomenological picture
}

Here, starting from the phenomenological model defined in Sec.~\ref{sec:model},
we shall deduce new results on the distribution $p(f)$ of  
the variable $f\equiv \ln Y_t-\ln Y_{t_0}$ 
(and on its moments) for $t,t_0,t-t_0\gg 1$,
up to one single constant for the moments of order larger than 2, 
and up to an
additional constant for the first moment.
Throughout, we shall aim at the accuracy 
${\cal O}\left(1/\sqrt{t_0},1/\sqrt{t},1/\sqrt{t-t_0}\right)$
for $p(f)$
and neglect higher powers of these expansion variables.

\subsection{\label{sec:distrib1}Effect of a fluctuation on $\ln Y$}

We first compute $\bar Y_t$, namely the variable
$Y_t$ in the mean-field approximation with the cutoff
in the tail. Using the definition of the variable $Y_t$ in Sec.~\ref{sec:var}
and using Eqs.~(\ref{eq:shape}),(\ref{eq:Xt}),
we find
\begin{equation}
\bar Y_t=\sqrt{t}\int_{-\infty}^{+\infty}dx\, \psi_{\bar X_t}(x,t)e^{x-2t}=-2\alpha
\left(1+\frac{C_{\bar Y}}{\sqrt{t}}\right),
\label{eq:Y0}
\end{equation}
at first order in $1/\sqrt{t}$.
$C_{\bar Y}$ is a constant literally equal to $C_{\bar X}-\frac{\beta\sqrt{\pi}}{2\alpha}$
in this calculation,
but there would also be other
contributions to $C_{\bar Y}$ 
that we cannot get from the large-$t$ asymptotic shape
of the front exhibited in Eq.~(\ref{eq:shape}).
We shall postpone the full calculation 
of $C_{\bar Y}$ to Sec.~\ref{sec:det}. 

It turns out that
the term of order $1/\sqrt{t}$ in Eq.~(\ref{eq:Y0})
generates ${\cal O}(1/t)$ contributions to the
distributions and to the moments that we shall address.
Hence it is enough for our purpose to keep no more than
the constant term, namely we write
\be
\bar Y_t\simeq -2\alpha.
\label{eq:Y00}
\ee
We perform a more
complete calculation in Appendix~\ref{sec:details},
keeping the subleading terms, 
in order to demonstrate that this
a priori approximation is indeed accurate enough.

Let us consider 
a fluctuation occurring at time $t_1\gg 1$ at a distance $\delta$ from the
tip of the deterministic front.
From Eq.~(\ref{eq:Xt}),
at time $t>t_1$ such that $t-t_1\gg1$, 
this fluctuation has developed its own front whose tip
sits at position 
\begin{equation}
\bar X_{\delta,t}=\bar X_{t_1}+\delta+\bar X_{t-t_1}
=\bar X_t+\delta-\frac32 \ln\frac{t_1(t-t_1)}{t}.
\label{eq:Xdelta}
\end{equation}
There would of course
be terms proportional to $1/\sqrt{t_1}$, $1/\sqrt{t}$ and $1/\sqrt{t-t_1}$
also here, but we again anticipate that they would eventually lead to
corrections of higher order to the quantities of interest.
We refer the reader to Appendix~A for the details.

The shape of the front generated by this fluctuation
will eventually
have the form $C\times \psi_{\bar X_{\delta,t}}(x,t-t_1)$,
where $C$ is a constant that we cannot determine since it is related
to some ``average'' shape
of the fluctuation.
With this extra fluctuation, $Y_t$ has the following expression:
\be
Y_t=\sqrt{t}\int_{-\infty}^{+\infty}dx\, \psi_{\bar X_t}(x,t)e^{x-2t}
+C\sqrt{t}\int_{-\infty}^{+\infty}dx\, \psi_{\bar X_{\delta,t}}(x,t-t_1)e^{x-2t}\ .
\ee
The first term is just $\bar Y_t$: We replace it by Eq.~(\ref{eq:Y00}).
The second term is integrated in the same way as the first one,
using the expression~(\ref{eq:Xdelta}) for $\bar X_{\delta,t}$.
We find
\be
Y_t=-2\alpha\left(1+C \frac{e^\delta}{t_1^{3/2}}\sqrt{\frac{t}{t-t_1}}\right).
\ee
Thus the forward shift in $\ln Y_t$ induced at time $t$ by such a fluctuation 
occurring at time $t_1$ reads
\be
\delta \ln Y_t=\ln Y_t-\ln\bar Y_t
=\ln\left(1+C \frac{e^\delta}{t_1^{3/2}}\sqrt{\frac{t}{t-t_1}}\right).
\label{eq:shift}
\ee
Note that in the asymptotic limit of interest, at first glance,
the nontrivial term in this expression seems to be of order $1/t_1^{3/2}$, thus,
if $t_1\sim t_0$, it is
smaller than our accuracy goal.
However, it is enhanced by the $e^{\delta}$ factor, 
which turns out to be large.

\subsection{Probability distribution and moments}

We may convert the conjectured probability of a forward fluctuation of size $\delta$
(Eq.~(\ref{eq:proba}))
into the probability distribution of the difference of $\ln Y$ between two times $t_0$ and
$t>t_0$
by simple changes of variables.
We first discuss the variable
\be
\delta f\equiv\delta \ln Y_t-\delta \ln Y_{t_0}.
\ee
The fundamental observation is that a fluctuation may essentially have 
two opposite effects on 
$\delta f\equiv \delta\ln Y_t-\delta\ln Y_{t_0}$.
If it occurs after time $t_0$, then it gives a positive contribution.
If instead it occurs before $t_0$, it generates a negative $\delta f$.
Now we observe that the difference between $\delta f$ and $f$ reads
$\ln {\bar Y_t}/{\bar Y_{t_0}}$, which is of order $1/\sqrt{t},1/\sqrt{t_0}$ and
thus, we may trade $\delta f$ for $f$ (see Appendix~\ref{sec:details} 
for more details).
\\

Let us first address the case in which the
fluctuation occurs between $t_0$ and $t$.
Using Eq.~(\ref{eq:shift}) together with the distribution~(\ref{eq:proba}),
the probability that the size of the shift in 
$\delta f$ induced by a fluctuation 
at time $t_1$ is less than some $F$ reads
\be
P(f<F;t_1)=C_1\left(
1-\frac{C}{t_1^{3/2}}\sqrt{\frac{t}{t-t_1}}\frac{e^{-F}}{1-e^{-F}}
\right).
\ee
We shall always assume that $F$ is finite, and the ordering $t,t-t_1,t_1\gg 1$.
The probability distribution of $f$ then reads
\be
p(f;t_1)=\frac{\partial P(f<F;t_1)}{\partial F}|_{F=f}=
\frac{CC_1}{t_1^{3/2}}\sqrt{\frac{t}{t-t_1}}\frac{e^{-f}}{\left(1-e^{-f}\right)^2}.
\label{eq:pf01}
\ee
The rate of the fluctuations is assumed constant in time, thus the distribution
of $f$ results from a simple integration over $t_1$
from $t_0$ to $t$
with uniform measure. It reads
\be
p(f)=2CC_1\sqrt{\frac{1}{t_0}-\frac{1}{t}}\frac{e^{-f}}{\left(1-e^{-f}\right)^2}
\ \ \text{for $f>0$}.
\label{eq:pf1}
\ee

Exactly in the same way, we may compute the probability distribution
of $f$ when the fluctuation occurs at a time smaller than $t_0$.
In this case, the effect on $f$ of a fluctuation of size $\delta$ reads
\be
f=\ln \frac{1+C \frac{e^\delta}{t_1^{3/2}}\sqrt{\frac{t}{t-t_1}}}
{1+C \frac{e^\delta}{t_1^{3/2}}\sqrt{\frac{t_0}{t_0-t_1}}}.
\ee
Using the same method, we find
\be
p(f)=2C C_1 \sqrt{\frac{1}{t_0}-\frac{1}{t}}\frac{e^{f}}{\left(1-e^{f}\right)^2}
\left(
1-\sqrt{\frac{1-e^f}{1+e^f}}
\right)
\ \ \text{for $f<0$}.
\label{eq:pf2}
\ee
The integral over $t_1$ 
which has to be performed to arrive at these expressions
is dominated by values of $t_1$ of the order of $t_0$.
This helps us to understand 
a posteriori why we were allowed to drop terms of order
$1/\sqrt{t}$ and $1/\sqrt{t_1}$ in Eqs.~(\ref{eq:Y0}) and~(\ref{eq:Xdelta}), 
respectively, although we were aiming at such accuracy for $p(f)$.

The probability distribution given in Eq.~(\ref{eq:pf1}) and~(\ref{eq:pf2})
cannot be normalized and the first moment $\langle f\rangle$ is also divergent. We
shall compute the latter separately in the next section.\\

An analytic continuation
of the generating function for the moments of $f$ can be obtained from Eq.~(\ref{eq:pf1})
and~(\ref{eq:pf2}) by a direct
calculation. We get
\be
\left\langle
e^{\nu f}
\right\rangle
=
2CC_1\sqrt{\frac{1}{t_0}-\frac{1}{t}}
\left\{
-\nu\psi(-\nu)+\nu\psi(\nu)+\sqrt{\pi}
\left[
\frac{\Gamma\left(\frac12+\frac{\nu}{2}\right)}
{\Gamma\left(\frac{\nu}{2}\right)}+
\frac{\Gamma\left(1+\frac{\nu}{2}\right)}
{\Gamma\left(\frac12+\frac{\nu}{2}\right)}
\right]
\right\},
\label{eq:moments}
\ee
keeping in mind that this formula can be used only for
moments of second order or higher.

The analytical structure of Eq.~(\ref{eq:moments}) is particularly
simple. There are poles on the positive real $\nu$-axis, fully 
contained in the first
term $-\nu\psi(-\nu)$: They correspond to positive values of $f$.
All the other poles, contained in the remaining terms, are located
on the real negative axis, and correspond to negative
values of $f$.
\\

A comment is in order on the conjectured probability 
distribution~(\ref{eq:proba}) of
the tip fluctuations that
we used in the above derivation.
Actually, we omitted a time-dependent 
Gaussian factor of the form $e^{-\delta^2/(4t_1)}$, 
which would cut off the exponential
distribution of $\delta$
at a distance $2\sqrt{t_1}$ ahead of the tip of the front, and thus modify
the distribution of $f$ for large positive $f$.
However, numerically, we do not find evidence for such a modification:
It seems that Eq.~(\ref{eq:pf1}) has a more general validity.
We do not have a good explanation for this surprising fact in the context
of our phenomenological model for fluctuations. But it turns out that a different
calculation of the positive $f$ fluctuations outlined in Appendix~\ref{sec:exact}
does not have such limitations.


\subsection{\label{sec:mu1}
Correction to the first moment of $f$ due to the fluctuations}

Since it is not possible to use Eq.~(\ref{eq:moments}) to get the first moment of $f$,
we shall arrive at its expression through a direct calculation.
We must take into account the expansion 
(keeping terms at least as large as $1/\sqrt{t}$, $1/\sqrt{t_0}$, $1/\sqrt{t-t_0}$)
of the density of particles in the deterministic
limit with a discreteness cutoff, and, in addition, 
the effect of the fluctuations
which intermittently speed up the evolution.
We have already guessed 
that there is an ${\cal O}(1/\sqrt{t})$ contribution to the deterministic
evolution (see Eq.~(\ref{eq:Y0})), but a full calculation will eventually
be needed.
Here, we shall simply denote by $C_{\bar Y}$ its coefficient.\\

The average of $f=\ln Y_t-\ln Y_{t_0}$ over realizations has thus a mean-field
contribution, and a contribution from the fluctuations which in turn
can be decomposed
in positive and negative contributions $\mu_1^+$ and $\mu_1^-$ respectively. We
shall evaluate the latter in this section. 

We write
\be
\mu_1=\langle \ln Y_t-\ln Y_{t_0} \rangle=C_{\bar Y}
\left(\frac{1}{\sqrt{t}}-\frac{1}{\sqrt{t_0}}\right)
+\mu_1^{+}-\mu_1^{-}.
\label{eq:c1}
\ee
Using Eq.~(\ref{eq:proba}) and Eq.~(\ref{eq:shift}), we get the
expression
\be
\mu_1^{+}=\int_{t_0^\prime}^{t}dt_1 \int_0^{+\infty} d\delta C_1 e^{-\delta}
\ln\left(1+C \frac{e^\delta}{t_1^{3/2}}\sqrt{\frac{t}{t-t_1}}\right)
\ee
for the positive part of the contribution at $t$
of the fluctuations, and
\be
\mu_1^{-}=\int_{t_0^\prime}^{t_0}dt_1 \int_0^{+\infty} d\delta C_1 e^{-\delta}
\ln\left(
{1+C \frac{e^\delta}{t_1^{3/2}}\sqrt{\frac{t_0}{t_0-t_1}}}\right)
\ee
subtracts the effect at $t_0$ of the fluctuations occurring at
$t_1<t_0$. 
We have introduced a time $t_0^\prime$ of order 1 as a lower bound
in these integrals in order to make these expressions finite.
The physical meaning of this cutoff is clear: Before $t_0^\prime$, there is no
mean-field regime because the whole system consists in a few particles
only.

Let us start with the computation of $\mu_1^+$.
It is useful to perform the change of variables
\be
\lambda=\frac{t_1}{t}\ ,\ \ u_\delta=e^{-\delta}\frac{t^{3/2}}{C}\lambda^{3/2}\sqrt{1-\lambda},
\label{eq:change}
\ee
which leads to the following expression of $\mu_1^{+}$:
\be
\mu_1^{+}=\frac{C C_1}{\sqrt{t}}\int_{\frac{t_0^\prime}{t}}^1 \frac{d\lambda}{\lambda^{3/2}}
\frac{1}{\sqrt{1-\lambda}}
\int_0^{u_0(\lambda)}du_\delta \ln\left(1+\frac{1}{u_\delta}\right),
\label{eq:mu1+}
\ee
where $u_0(\lambda)=\frac{t^{3/2}}{C}\lambda^{3/2}\sqrt{1-\lambda}$.
$u_0$ is large compared to 1, except when $1-\lambda$ is of order $1/t^3$.
But the contribution of the region $[1-1/t^3,1]$ in 
the $\lambda$-integration
is smaller than $\sim 1/t^{3/2}$, and hence negligible.
So we may always assume $u_0\gg 1$.

The integral over $u_\delta$ is performed
analytically, and the large-$u_0$ limit may eventually be taken:
\be
\int_0^{u_0}du_\delta\ln\left(1+\frac{1}{u_{\delta}}\right)=(1+u_0)\ln(1+u_0)-u_0\ln u_0
\underset{u_0\gg 1}{\sim} \ln u_0.
\ee
The remainder
reads
\be
\mu_1^+=\frac{CC_1}{\sqrt{t}}\left[
\ln \left(\frac{t^{3/2}}{C}\right)
{\cal I}_0
+{\cal I}_1
\right],
\label{eq:mu1+I0I1}
\ee
where
\be
{\cal I}_0=\int_{t_0^\prime/t}^1 d\lambda\lambda^{-3/2}(1-\lambda)^{-1/2},
\ \
{\cal I}_1=\int_{t_0^\prime/t}^1 d\lambda\lambda^{-3/2}(1-\lambda)^{-1/2}
\ln \left(\lambda^{3/2}\sqrt{1-\lambda}\right).
\label{eq:defI0}
\ee
${\cal I}_0$ and ${\cal I}_1$ can be performed
with the help of the change of variable $\lambda=\sin^2\theta$:
\be
{\cal I}_0=2\int_{\arcsin\sqrt{\frac{t_0^\prime}{t}}}^{\frac{\pi}{2}}\frac{d\theta}{\sin^2\theta}
=2 \cot\left(\arcsin \sqrt{\frac{t_0^\prime}{t}}\right)
=2\sqrt{\frac{t}{t_0^\prime}-1},
\ee
while for ${\cal I}_1$, a further integration by parts is needed to
get rid of the log.
We eventually arrive at the following exact expression for~(\ref{eq:mu1+I0I1}):
\be
\mu_1^{+}=2CC_1\left\{\sqrt{\frac{1}{t_0^\prime}-\frac{1}{t}}
\left[
\ln \left(
\frac{t_0^{\prime 3/2}}{C}\sqrt{1-\frac{t_0^\prime}{t}}
\right)+3
\right]
-\frac{4}{\sqrt{t}}\arccos\sqrt{\frac{t_0^\prime}{t}}
\right\}.
\label{eq:c1+}
\ee
The term $\mu_1^-$ is the same as the term $\mu_1^+$
except for the replacement $t\rightarrow t_0$.

Since we are neglecting terms of relative
order $t_0/t$, $t_0^\prime/t_0$ and higher, 
we may expand the expressions for $\mu_1^+$ and $\mu_1^-$.
The difference $\mu_1^+-\mu_1^-$
then reads
\be
\mu_1^+-\mu_1^-=4\pi CC_1\left(\frac{1}{\sqrt{t_0}}-\frac{1}{\sqrt{t}}\right).
\ee
Equation~(\ref{eq:c1}) eventually leads to the following expression for $\mu_1$:
\be
\mu_1=(4\pi C C_1-C_{\bar Y})\left(\frac{1}{\sqrt{t_0}}-\frac{1}{\sqrt{t}}\right).
\label{eq:mu1}
\ee

We note a very important property of this result: It does not depend
on $t_0^\prime$. If it did, then we would loose predictivity because
$t_0^\prime$ is the arbitrary time after which we declare that the 
fluctuations are small enough for our calculation to apply.
(This would not be true at the next order in $1/\sqrt{t}$, $1/\sqrt{t_0}$).


\section{\label{sec:det}Deterministic calculations}

In this section, we first review the Ebert-Van Saarloos method \cite{EVS:2000}
to compute the order $1/\sqrt{t}$ correction to 
the mean position of the rightmost particle in the BRW
$\langle X_t\rangle$. We
extend the method to 
the position of the right boundary in the deterministic
model with discreteness cutoffs
$\bar X_t$, and eventually
adapt it to $\langle \ln Y_t\rangle$.

The calculations presented here will enable us to determine the remaining unknown
constants, namely $C_{\bar X}$ (see Eq.~(\ref{eq:Xt})),
$C_{\bar Y}$ (Eq.~(\ref{eq:Y0})), and $CC_1$.
The latter two constants
appear in particular
in Eqs.~(\ref{eq:pf1}), (\ref{eq:pf2}), (\ref{eq:moments}) and~(\ref{eq:mu1}).

\subsection{\label{sec:noncritical}
Ebert-Van Saarloos calculation and its extension}

The original calculation of Ebert and Van Saarloos aimed at finding
properties of the solutions to the FKPP equation
\be
\partial_t \phi(t,x)=\partial_x^2\phi(t,x)+\phi(t,x)-\phi^2(t,x)
\label{eq:FKPPphi}
\ee
for $\phi\ll 1$, with a steep enough initial condition,
e.g. $\phi(t=0,x)=\theta(-x)$. This equation is actually the same as Eq.~(\ref{eq:FKPP}),
with the correspondence $\phi(t,x)=1-Q(t,x)$.
The nonlinearity can essentially be viewed as a moving absorptive boundary
 on a linear
partial differential equation, the position of the boundary being set
in such a way that
$\phi$ has a maximum
at a fixed height.

To determine the value of the constants $C_{\bar X}$ and $C_{\bar Y}$,
we need to address a branching diffusion equation with a nonlinearity
that forces $\phi$ to go to zero over a distance of order~1 at the right of
the point at which $\phi(t,x)=1$, and which therefore acts as a tip cutoff. 
In terms of a smooth equation, we may
write, for example,
\be
\partial_t \phi(t,x)=\partial_x^2\phi(t,x)+\frac{\phi^2(t,x)}{1+\phi(t,x)}
\label{eq:tipcutoffsmooth}
\ee
and study the properties of the solutions to this equation
in the region $\phi\gg 1$.

In both cases, the equation can be linearized in the respective domain
of interest, and one gets
\be
\partial_t \phi(t,x)=\partial_x^2\phi(t,x)+\phi(t,x).
\ee
We shall assume that the nonlinear term is equivalent to a right-moving 
absorptive boundary at the accuracy at which we want to address the problem.
(This assumption was better motivated by Ebert and Van Saarloos in their discussion
of what they call the ``interior expansion''~\cite{EVS:2000}).
In the first case, we study the function $\phi$ to the right of the boundary;
in the second case, we study the function to the left.

\paragraph{Solution to the linearized equation with an appropriate boundary condition.}

Near the boundary, at large times, the function $\phi$ reads
\be
\phi(t,x)\sim (\alpha\xi+\beta)e^{-\xi},
\label{eq:phiinterior}
\ee
where $\xi=x-\text{[position of the cutoff]}\sim x-2t+\frac32\ln t+\dots$
and this is valid for $1\ll \xi\ll \sqrt{t}$.
According to Ebert-Van Saarloos~\cite{EVS:2000}, 
the large-$t$ corrections to this shape
are of the form $1/t$ (there is no term of order $1/\sqrt{t}$).
All these features should not depend on whether we address
Eq.~(\ref{eq:FKPPphi}) or Eq.~(\ref{eq:tipcutoffsmooth}) above,
except for the signs of $\alpha$ and $\xi$.

We write
\be
\phi(t,x)=e^{-\xi-z}g(t,z)\ ,
\ee
where $z=\frac{\xi^2}{4t}$, and the following ansatz
are taken:
\be
\begin{split}
\xi&=x-2t+\frac32\ln t+\frac{2c}{\sqrt{t}}+\cdots\\
g(t,z)&=\sqrt{t}\, g_{-\frac12}(z)+g_0(z)+\cdots
\end{split}
\label{eq:ansatzg}
\ee
The variable $\xi$ may be positive in the linear domain
(it is the case for the usual Ebert-Van Saarloos solution)
or negative: Therefore, we write $\xi=\varepsilon\sqrt{4t\,z}$, 
where $\varepsilon=\pm 1$.
The ansatz for the front position contained in 
$\xi$ already incorporates the two
known \cite{B:1983} dominant terms at large $t$, 
namely $2t-\frac32\ln t$.
The $-\frac{2c}{\sqrt{t}}$ term was new
in Ref.~\cite{EVS:2000}.

Thanks to these ansatz, the original equation splits into a hierarchy of 
equations for the functions $g$. The first two equations of this set read
\be
\begin{split}
&z g_{-\frac12}^{\prime\prime}+\left(\frac12-z\right)
g_{-\frac12}^\prime+\frac12 g_{-\frac12}=0\ ,\\
&z g_{0}^{\prime\prime}+\left(\frac12-z\right)
g_{0}^\prime+g_{0}= c g_{-\frac12}-\varepsilon\frac32\sqrt{z}
(g_{-\frac12}-g^\prime_{-\frac12})\ .
\end{split}
\label{eq:hierarchy}
\ee
The first equation of the hierarchy is the Kummer equation
\be
z\frac{d^2w}{dz^2}+(b-z)\frac{dw}{dz}-aw=0
\ee
with $w=g_{-\frac12}$, $a=-\frac12$  and $b=\frac12$.
Two independent solutions are, for example, the two Kummer functions
(or ${}_1F_1$ hypergeometric functions)
\be
M(a,b,z)\ \
\text{and}\ \
z^{1-b}M(a-b+1,2-b,z)
\ee
namely, in our case,
\be
M\left(-\frac12,\frac12,z\right)\ \ \text{and}\ \ \sqrt{z}\,
M\left(0,\frac32,z\right).
\ee
The latter
is just the elementary function
$\sqrt{z}$, while the former
diverges like $-e^{z}/(2z)$ for large $z$, and has thus to be discarded.
Hence the solution reads
\be
g_{-\frac12}=2\alpha^\prime\sqrt{z}
\label{eq:g-12sol}
\ee
where $\alpha^\prime$ is a constant, arbitrary at this stage.

As for the second equation in Eq.~(\ref{eq:hierarchy})
whose solution is the function $g_0$,
it is an inhomogeneous Kummer differential equation.
A basis for the solutions of the homogeneous part 
is, for example, the set of the two functions
\be
M\left(-1,\frac12,z\right)=1-2z\ \ \text{and}\ \ \sqrt{z}\,M\left(-\frac12,\frac32,z\right).
\ee
We need to find a particular solution of the full equation.
We define
$y\equiv\sqrt{z}$;
The equation for $g_0$ then reads
\be
\frac{d^2 g_0}{dy^2}
-2y\frac{dg_0}{dy}+4g_0
=
8\alpha^\prime\left(
-\frac32\varepsilon y^2
+cy
+\frac34\varepsilon
\right)
\label{eq:diffg0}
\ee
and we may look for solutions in terms of a series:
\be
g_0(y)=\sum_{k=0}^{+\infty}a_k y^k\ .
\ee
Inserting this expression into the
differential equation~(\ref{eq:diffg0}),
we get the following relations between the coefficients of the series:
\be
a_{k+2}=\frac{2(k-2)}{(k+1)(k+2)}a_k\ \text{for $k\geq 3$}\ ,
\ \ a_2=-2a_0+3{\alpha^\prime}\varepsilon\ ,
\ \ a_3=-\frac{a_1}{3}+\frac{4{\alpha^\prime} c}{3}\ ,
\ \ a_4=-{\alpha^\prime}\varepsilon\ .
\ee
The free parameters are $a_0$ and $a_1$.
We may choose them in such a way that $a_{2,3}=0$:
We therefore set $a_0=\frac32{\alpha^\prime}\varepsilon$ and $a_1=4{\alpha^\prime} c$.
Then
\be
a_{2n}=-\frac32\sqrt{\pi}{\alpha^\prime}\varepsilon\frac{\Gamma(n-1)}{\Gamma(n+1)\Gamma(n+1/2)}
\ \text{for $n\geq 2$},
\ \text{$a_{2}=0$, and
$a_{2n+1}=0$ for $n\geq 1$},
\ee
where we used the duplication formula
$\Gamma(2n+1)=\frac{2^{2n}}{\sqrt{\pi}}
\Gamma(n+\frac12)\Gamma(n+1)$.
Switching back to the variable $z$, the final expression for the
particular solution reads
\be
g_0^{sp}(z)=\frac32{\alpha^\prime}\varepsilon+4{\alpha^\prime} c \sqrt{z}
-\frac{3{\alpha^\prime}\varepsilon}{2}F_2(z)
\ \ \text{where}\ \
F_2(z)=\sqrt{\pi}
\sum_{n=2}^{\infty}\frac{\Gamma(n-1)}{\Gamma(n+1/2)\Gamma(n+1)}z^n\ ,
\ee
which, except for the sign factors~$\varepsilon$, is the Ebert-Van Saarloos result \cite{EVS:2000}.
Following again Ref.~\cite{EVS:2000}, we write the solution for $g_0$ as 
\be
g_0(z)=\frac32{\alpha^\prime}\varepsilon+4{\alpha^\prime} c \sqrt{z}
-\frac{3{\alpha^\prime}\varepsilon}{2}F_2(z)+k_0(1-2z)
+l_0\sqrt{z}\,M\left(
-\frac12,\frac32,z
\right)
\label{eq:g0sol}
\ee
and inserting~(\ref{eq:g0sol}) together with~(\ref{eq:g-12sol})
into~(\ref{eq:ansatzg}),(\ref{eq:phiinterior}), we would
get the expression of $\phi$ up to the constants $\alpha^\prime,c,k_0,l_0$.
We are now going to determine them from a matching procedure.

\paragraph{Matching conditions.}

We now match with the shape of the so-called ``interior'' region at $z\ll 1$.
This means that $\phi$ just obtained
should have the same small-$z$ expansion as the limiting form of
$\phi$ in Eq.~(\ref{eq:phiinterior}).
Hence we need to impose
\be
g_{-\frac12}(z)\underset{z\ll 1}\sim 2\alpha\sqrt{z}
\ \ \text{and} \ \
g_0(z)\underset{z\ll 1}\sim\beta+{\cal O}(z).
\ee
The first constraint is solved by setting $\alpha^\prime=\alpha$. As for
the second one, it means 
in particular that there should be no term proportional to $\sqrt{z}$ in $g_0(z)$.
This requirement leads to the equations
\be
\frac{3\alpha\varepsilon}{2}+k_0=\beta\ ,\ \
4\alpha c+{l_0}=0.
\label{eq:require}
\ee

Now we must also check the behavior at $z\rightarrow+\infty$.
We need the expansion of the functions
$M$ and $F_2$ for $z\rightarrow\infty$.
Let us start with $M$. We shall use the integral representation
\be
M(a,b,z)=\frac{\Gamma(b)}{\Gamma(a)\Gamma(b-a)}
\int_0^1 du\,e^{zu} u^{a-1}(1-u)^{b-a-1}.
\ee
We change the variable for $u$ to $1-u$ in the integral,
and we expand the $(1-u)^{a-1}$ factor near $u=0$:
\be
M(a,b,z)=
\frac{\Gamma(b)}{\Gamma(a)\Gamma(b-a)}
e^z\int_0^1 du\, e^{-zu}u^{b-a-1}\sum_{k=0}^{+\infty}
\frac{\Gamma(1-a+k)}{\Gamma(1-a)\Gamma(1+k)}u^k.
\ee
We then notice that we may extend the integral to $+\infty$ without
adding exponentially-enhanced terms.
Finally, we perform the remaining integration over $u$.
The result reads
\be
\begin{split}
M(a,b,z)&=e^zz^{a-b}\frac{\Gamma(b)}{\Gamma(a)}
\sum_{k=0}^{+\infty}\frac{z^{-k}}{\Gamma(1+k)}
\frac{\Gamma(1-a+k)\Gamma(b-a+k)}{\Gamma(1-a)\Gamma(b-a)}
+o(e^z)\\
&=e^z z^{a-b}\frac{\Gamma(b)}{\Gamma(a)}{}_2F_0(1-a,b-a;1/z)+o(e^z).
\end{split}
\ee
Setting $a=-1/2$ and $b=3/2$, we write
\be
\sqrt{z}M\left(-\frac12,\frac32,z\right)
\sim -\frac14 e^{z}z^{-3/2}
\sum_{k=0}^{+\infty}\frac{\Gamma(\frac32+k)}{\Gamma(\frac32)}(1+k)z^{-k}
=-\frac14 e^z z^{-3/2}{}_2F_0\left(\frac32,2;;\frac{1}{z}\right).
\ee
We now turn to $F_2$.
We write the following integral representation:
\be
\frac12 F_2(z)=\lim_{\eta\rightarrow 0} \left[
\int_0^1 du\, e^{zu}u^{-2+\eta}\sqrt{1-u}
-
\frac{\sqrt{\pi}}{2}\frac{\Gamma(\eta-1)}
{\Gamma\left(\eta+\frac12\right)}-
\frac{\sqrt{\pi}}{2}\frac{\Gamma(\eta)}
{\Gamma\left(\eta+\frac32\right)}z
\right].
\ee
This representation may be checked by expanding the exponential
in the integral and performing the integration over $u$.
For large $z$, the two rightmost terms do not play any role since
they are not exponentially enhanced.
We may now treat the first term exactly in the same way as
in the case of the Kummer function $M$.
After taking the $\eta\rightarrow 0$ limit, which is
finite once all non-exponentially enhanced terms have been discarded,
we get
\be
F_2\sim 2e^{z}z^{-3/2}\sum_{k=0}^{+\infty}
\Gamma\left(k+\frac32\right)(k+1)z^{-k}.
\ee
Up to an overall constant, all terms are identical to the ones in
the expansion of the $M$ function.

Requiring the cancellation of these exponentially-enhanced
terms in the expression (\ref{eq:g0sol})
for $g_0$ leads to the equation
\be
\frac32\alpha\varepsilon\sqrt{\pi}+\frac{l_0}{4}=0\ .
\ee
Using this equation and the second equation
in~(\ref{eq:require}), one determines the value of $c$:
\be
c=\frac32\varepsilon\sqrt{\pi}\ .
\ee
Hence this constant is {\em positive} for the Ebert-Van Saarloos solution
of the FKPP equation, but is {\em negative} when one computes the position of the tip
of a front with a discretness cutoff.

\paragraph{Matched solution.}

All in all, we get
\be
\phi(t,x)=e^{-\xi-z}\left\{
\alpha\xi+\beta
+\left(
3\alpha\varepsilon-2\beta
\right)z
+6\alpha\varepsilon\sqrt{\pi z}
\left[1-M\left(-\frac12,\frac32,z\right)\right]
-\frac{3\alpha\varepsilon}{2}F_2(z)
\right\}
\label{eq:phinoncrit}
\ee
with
\be
\xi=x-2t+\frac32\ln t+3\varepsilon\sqrt{\frac{\pi}{t}}.
\ee
The first two terms in $\phi$, namely $e^{-\xi-z}(\alpha\xi+\beta)$, give
back Eq.~(\ref{eq:shape}). The next terms are finite-time corrections.

Identifying $\xi$ with $x-\langle X_t\rangle$ and
setting $\varepsilon=+1$, we recover the value of 
\be
C_{X}=-3\sqrt{\pi}
\label{eq:CX}
\ee
already derived by Ebert and Van Saarloos (see Eq.~(\ref{eq:FKPPpos})).
With $\xi=x-\bar X_t$ and $\varepsilon=-1$, we read off this formula the 
value of the constant
\be
C_{\bar X}=3\sqrt{\pi}.
\label{eq:CMFX}
\ee
We can also deduce the value of $C_{\bar Y}$ by using the
definition of the variable $Y_t$ 
given in Sec.~\ref{sec:var}
and the shape of the mean-field particle
distribution~(\ref{eq:phinoncrit}):
\be
\bar Y_t=\sqrt{t}\int_{-\infty}^{+\infty}dx\,
\phi(t,x)e^{x-2t},
\ee
which, after replacement by the expression~(\ref{eq:phinoncrit}) and
setting $\varepsilon=-1$, becomes
\begin{multline}
\bar Y_t=e^{3\sqrt{\pi/t}}
\bigg\{
-2\alpha-\frac{1}{\sqrt{t}}
\int_0^{+\infty}\frac{dz}{\sqrt{z}}e^{-z}
\bigg[
\beta(1-2z)\\
-3\alpha z-6\alpha\sqrt{\pi z}
\left\{1-M\left(-\frac12,\frac32,z\right)\right\}
+\frac{3\alpha}{2}F_2(z)
\bigg]
\bigg\}.
\end{multline}
The term proportional to $\beta$ is zero after the integration,
and the other terms give numerical constants.
We finally find, at order $1/\sqrt{t}$,
\be
{\ln \bar Y_t}=\ln(-2\alpha)+\frac{C_{\bar Y}}{\sqrt{t}}\ ,
\ \ \text{with}\ \ C_{\bar Y}=\frac 32{\sqrt{\pi}}.
\label{eq:CMFY}
\ee


\subsection{\label{sec:critical}
Solution of the deterministic FKPP equation with the critical initial condition
}

Let us consider a generating function of the moments of the $\tilde Y_t$ variable:
\be
G_t(x)=\left\langle e^{-\tilde Y_t e^{-x}}\right\rangle=
\left\langle \prod_{i=1}^{N(t)} e^{-e^{-(x-x_i(t))}}\right\rangle.
\ee
Defining $f(x)=e^{-e^{-x}}$, $G_t(x)$ has exactly the form shown in Eq.~(\ref{eq:G}) and
thus
$\phi(t,x)\equiv 1-G_t(x)$ 
solves the FKPP equation~(\ref{eq:FKPPphi}),
$\partial_t\phi=\partial_x^2\phi +\phi-\phi^2$.
If the initial condition
for the underlying
branching random walk is a single particle at position $x=0$,
\be
\phi(t=0,x)=1-e^{-e^{-x}}
\ee
and then, the position of the FKPP traveling wave is given by Eq.~(\ref{eq:FKPPposcrit}).
In this section, we shall address this case using the Ebert-Van Saarloos method in order
to obtain the $1/\sqrt{t}$ correction to the latter and some analytic features of $\phi$.
Indeed, from the expression of
$\phi$, we may in principle compute the moments of $\ln\tilde Y_t$, using
the identity 
\be
\left\langle \tilde Y_t^\nu\right\rangle=
-\frac{1}{\Gamma(1-\nu)}
\int_{-\infty}^{+\infty}dx\,e^{\nu x}\frac{\partial\phi(t,x)}{\partial x}.
\label{eq:MomentsFromPhi}
\ee

\paragraph{General solution of the linearized equation in a moving frame.}

Following Ebert-Van Saarloos, we define
\be
\xi=x-2t-\chi_t\ \ \text{and}\ \
\phi(t,x)=e^{-\xi}\psi(t,\xi).
\label{eq:defpsi}
\ee
The linearized FKPP equation for $\psi$ reads
\be
\partial_t\psi(t,\xi)=\partial_\xi^2\psi(t,\xi)+\dot \chi_t(\partial_\xi-1)\psi(t,\xi).
\ee
Next, we take the ansatz
$\chi_t=-\frac12\ln t-\frac{2 c}{\sqrt{t}}$,
and introduce the variable $z=\frac{\xi^2}{4t}$.
The function $g(t,z)$ is $\psi(t,\xi)$ expressed with the help of $z$,
and we look for solutions
in the form
\be
g(t,z)=\sqrt{t}\,g_{-\frac12}(z)+g_0(z).
\label{eq:defg}
\ee
We are led to the following hierarchical set of equations (compare to Eq.~(\ref{eq:hierarchy})):
\be
\begin{split}
&zg_{-\frac12}^{\prime\prime}(z)+\left(z+\frac12\right)g_{-\frac12}^\prime(z)=0\\
&zg_0^{\prime\prime}(z)+\left(z+\frac12\right)g_0^\prime(z)+\frac12 g_0(z)
=c\, g_{-\frac12}(z)+\frac12\sqrt{z}\,
g_{-\frac12}^\prime(z).
\end{split}
\ee
The solution reads
\be
\begin{split}
&g_{-\frac12}(z)=b+a\sqrt{\pi}\erf(\sqrt{z})\\
&g_0(z)=
\left[\frac{c_1\sqrt{\pi}}{2}e^{-z}\erfi(\sqrt{z})+c_2e^{-z}\right]
+2c\left[b+a\sqrt{\pi}\erf(\sqrt{z})\right]+
az e^{-z}{}_2F_2(1,1;{\scriptstyle \frac32},2;z)
\end{split}
\label{eq:solgen}
\ee
where $\erf,\erfi$ are the error functions defined by
\be
\erf(x)=\frac{2}{\sqrt{\pi}}\int_0^x dt\, e^{-t^2},\ \erfi(x)=-i \erf(ix),
\ee
and $a,b,c_1,c_2$ are integration constants to be determined.
The terms in the first
square brackets in Eq.~(\ref{eq:solgen})
correspond to the general solution of the
homogeneous equation for $g_0$, 
while the next two terms represent a particular solution of the
full equation as can easily be checked.

\paragraph{Matching conditions.}

Because of the initial condition, 
the tail of the front at $\xi\rightarrow \infty$
has the exact shape
\be
\phi(t,x\gg 2t\gg 1)=e^{-(x-2t)}
\ee
at any time. In particular, there is no overall constant.
Comparing to Eqs.~(\ref{eq:defpsi}),(\ref{eq:defg}), this condition means that
\be
g(t\rightarrow\infty,z\rightarrow\infty)=\sqrt{t}+2c+{\cal O}(1/\sqrt{t}).
\ee
Let us expand our solution~(\ref{eq:solgen}) for $g(t,z)$ for large $t,z$:
\begin{multline}
g(t\rightarrow\infty,z\rightarrow \infty)=
\sqrt{t}\left[b+a\sqrt{\pi}+{\cal O}(e^{-z})\right]
+2c(b+a\sqrt{\pi})\\
+\frac12(c_1+a\sqrt{\pi})\left(
\frac{1}{\sqrt{z}}+\frac{1}{2z^{3/2}}+\cdots
\right)
+{\cal O}(e^{-z}).
\end{multline}
The identification with the expected asymptotic form leads to the conditions:
\be
b+a\sqrt{\pi}=1,\ {2c}(b+a\sqrt{\pi})=2c.
\ee
The second condition is trivial once the first one is satisfied.

We also impose that all terms that are not exponentially suppressed cancel,
which
is realized by setting
\be
c_1+a\sqrt{\pi}=0.
\ee

We turn to the limit $z\rightarrow 0$.
The condition~(\ref{eq:phiinterior}) reads, in terms of the $g$-function
\be
g(t\rightarrow\infty,z\rightarrow 0)=\alpha\sqrt{tz}+\beta
\label{eq:match0}
\ee
which in particular forbids constant terms
and terms proportional to $\sqrt{z}$.
Since the small-$z$ expansion of our solution reads
\be
g(t\rightarrow\infty,z\rightarrow0)=\sqrt{t}\left[b+2a\sqrt{z}+{\cal O}(z)\right]\\
+2cb+c_2+(4ac+c_1)\sqrt{z}+{\cal O}(z)
\ee
we see that $b$ needs to be set to 0 and $c=-c_1/(4a)$.

Putting everything together, we find that all constraints are solved by the choice
\be
a=\frac{1}{\sqrt{\pi}},\ b=0,\ c=\frac{\sqrt{\pi}}{4},\ 
c_1=-1,\ c_2=\beta.
\ee
Note that the coefficient
$\alpha$ in Eq.~(\ref{eq:match0}) 
is determined to be $\alpha=2/\sqrt{\pi}$, while in the
noncritical case, it is a free parameter.

\paragraph{Matched solution.}

All in all, our solution reads
\be
\phi(t,x)=
e^{-\xi}\bigg[
\sqrt{t}\,
\erf\left(\sqrt{z}\right)
+e^{-z}
\left\{
\beta 
+\frac{z}{\sqrt{\pi}}\,{}_2F_2(1,1;{\scriptstyle\frac32},2;z)
+\frac{\sqrt{\pi}}{2}
\left[e^z\erf(\sqrt{z})
-\erfi(\sqrt{z})\right]
\right\}
\bigg],
\label{eq:phi}
\ee
with
\be
\xi=x-2t+\frac{1}{2}\ln t+\frac{\sqrt{\pi}}{2}
\frac{1}{\sqrt{t}}\,,\ \
z=\frac{\xi^2}{{4t}}\,.
\ee
The $1/\sqrt{t}$ term is identical to the one in the ``pushed front'' calculation
of Ref.~\cite{EVS:2000}, see Appendix~G, Eq.~(G18) therein, 
although the front solution chosen in
that work is different, see Eq.~(G7).

We can now deduce from this calculation the average value $\mu_1$
of $\ln Y_t-\ln Y_{t_0}$ by
expanding the exact formula Eq.~(\ref{eq:MomentsFromPhi}) in powers of $\nu$ and
keeping the coefficient of the term of order $\nu$:
\be
\langle\ln \tilde Y_t\rangle
=\langle\ln Y_t\rangle+2t-\frac12\ln t=-\psi(1)-\int_{-\infty}^{+\infty}dx\,x
\frac{\partial}{\partial x}\phi(t,x).
\label{eq:logYfromphi}
\ee
We find
\be
\mu_1=\langle\ln Y_t-\ln Y_{t_0}\rangle=\frac{\sqrt{\pi}}{2}\left(\frac{1}{\sqrt{t_0}}
-\frac{1}{\sqrt{t}}\right).
\ee
Identifying the latter equation to Eq.~(\ref{eq:mu1}) and taking into account
the value of $C_{\bar Y}$ already computed in Eq.~(\ref{eq:CMFY}), 
we finally obtain a determination of $CC_1$:
\be
CC_1=\frac{1}{2\sqrt{\pi}}.
\label{eq:CC1}
\ee


\section{\label{sec:numerics}Complete results and numerical checks}

Since the new results we have obtained rely in an essential way
on a model for fluctuations and hence on a set of conjectures, we need to
check them with the help of numerical simulations in order to
get confidence in the validity of our picture.
In the first part of this section, we shall list the formulas
we have obtained but extending them to more general BRW models.
Then, we define a model that is convenient for numerical implementation
in Sec.~\ref{sec:modelnum}, and we test our results  against
numerical simulations 
of this particular model 
in Secs.~\ref{sec:checkdet} and~\ref{sec:checkstat}.

\subsection{Parameter-free predictions for a general branching diffusion}

We now extend our results to general branching diffusion kernels.
In the continuous case, we write
the equation for the average
particle density as
\be
\partial_t{\langle n(t,x)\rangle}=
\chi(-\partial_x)\langle n(t,x)\rangle
\ee
where $\chi(-\partial_x)$ is the operator that represents the branching diffusion.
The eigenfunctions are the exponential functions $e^{-\gamma x}$, and the corresponding
eigenvalues are $\chi(\gamma)$.
In the case discussed in the previous sections, $\chi(-\partial_x)=\partial_x^2+1$
and $\chi(\gamma)=\gamma^2+1$.
We introduce
$\gamma_0$ which solves $\chi^\prime(\gamma_0)=\chi(\gamma_0)/\gamma_0$.
Then in the case studied so far,
$\gamma_0=1$ and $\chi(\gamma_0)=\chi^\prime(\gamma_0)=\chi^{\prime\prime}(\gamma_0)=2$.

We can also address the discrete time and space case, which is
useful in particular for numerical simulations.
We write
\be
\frac{\langle n(t+\Delta t,x)\rangle-\langle n(t,x)\rangle}
{\Delta t}=\chi(-\delta_x)\langle n(t,x)\rangle
\ee
where now $\delta_x$ is some finite difference operator, such as
\be
\delta_x f(x)=\frac{f(x+\Delta x)-f(x)}{\Delta x}.
\ee
In this case,
$t$ and $x$ take their values on lattices of respective 
spacing $\Delta t$ and $\Delta x$.
Again, the eigenfunctions of the kernel are the exponential functions.

The generalization of our previous results 
to an arbitrary BRW relies on the fact that at large times,
the ``wave number'' $\gamma_0$ dominates and the kernel eigenvalue
$\chi(\gamma)$ may be expanded to second order around $\gamma_0$ \cite{VS:2003}.
We then essentially use dimensional analysis
to put in the appropriate process-dependent factors. 
We list here the generalized expressions
without detailed justifications.

With the general kernel, the FKPP front position reads (see Eq.~(\ref{eq:FKPPpos}))
\be
\langle X_t\rangle=\chi'(\gamma_0)t-\frac{3}{2\gamma_0}\ln t+\text{const}
-\frac{3}{\gamma_0^2}\sqrt{\frac{2\pi}{\chi^{\prime\prime}(\gamma_0)}}\frac{1}{\sqrt{t}}+\cdots
\ee
The position of the tip of the front in the mean-field model with a discreteness cutoff
reads instead
\be
\bar X_t=\chi'(\gamma_0)t-\frac{3}{2\gamma_0}\ln t+\text{const}
+\frac{3}{\gamma_0^2}\sqrt{\frac{2\pi}{\chi^{\prime\prime}(\gamma_0)}}\frac{1}{\sqrt{t}}+\cdots
\ee
This expression generalizes Eq.~(\ref{eq:Xt}) with $C_X$ computed in Sec.~\ref{sec:noncritical}
(see Eq.~(\ref{eq:CMFX})).

The relevant variable that characterizes the fluctuations of the
position of the bulk of the particles
is $\frac{1}{\gamma_0}\ln Y_t$. We have computed its value in the deterministic
model with a tip cutoff:
\be
\frac{1}{\gamma_0}{\ln \bar Y_t}=
\text{const}+\frac{3}{2\gamma_0^2}\sqrt{\frac{2\pi}{\chi^{\prime\prime}(\gamma_0)}}
\frac{1}{\sqrt{t}}.
\label{eq:logYdettip}
\ee
This equation generalizes Eq.~(\ref{eq:CMFY}).

The stochasticity that we found tractable analytically is related to
the fluctuations
of the difference of this variable at two distinct large times
$t_0$ and $t$:
\be
f=\frac{1}{\gamma_0}
\left(
\ln Y_t-\ln Y_{t_0}
\right).
\ee
Its first moment reads
\be
\mu_1=\langle f \rangle=\frac{1}{2\gamma_0^2}\sqrt{\frac{2\pi}{\chi^{\prime\prime}(\gamma_0)}}
\left(
\frac{1}{\sqrt{t_0}}-\frac{1}{\sqrt{t}}
\right).
\label{eq:mu1fulldet}
\ee
The probability distribution of the fluctuations reads
\be
p(f)=
\begin{cases}
\sqrt{\frac{2}{\pi\chi^{\prime\prime}(\gamma_0)}}
\sqrt{\frac{1}{t_0}-\frac{1}{t}}
\frac{e^{-\gamma_0 f}}{\left(1-e^{-\gamma_0 f}\right)^2}
& \text{if $f>0$},\\
\sqrt{\frac{2}{\pi\chi^{\prime\prime}(\gamma_0)}}
\sqrt{\frac{1}{t_0}-\frac{1}{t}}
\frac{e^{\gamma_0f}}{\left(1-e^{\gamma_0f}\right)^2}\left[
1-\sqrt{\frac{1-e^{\gamma_0f}}{1+e^{\gamma_0f}}}
\right]& \text{if $f<0$}.
\end{cases}
\label{eq:pf}
\ee
This formula is the generalized form of Eqs.~(\ref{eq:pf1}) and~(\ref{eq:pf2}).
A generating function of the moments of order larger than~2
can be written as
\be
\left\langle
e^{\gamma_0\nu f}
\right\rangle
=\frac{1}{\gamma_0}
\sqrt{\frac{2}{\pi\chi^{\prime\prime}(\gamma_0)}}
\sqrt{\frac{1}{t_0}-\frac{1}{t}}
\left\{
-\nu\psi(-\nu)+\nu\psi(\nu)+\sqrt{\pi}
\left[
\frac{\Gamma\left(\frac12+\frac{\nu}{2}\right)}
{\Gamma\left(\frac{\nu}{2}\right)}+
\frac{\Gamma\left(1+\frac{\nu}{2}\right)}
{\Gamma\left(\frac12+\frac{\nu}{2}\right)}
\right]
\right\}.
\ee
For example, expanding this generating function,
we find that the moments of order $k\geq 2$ read
\be
\mu_k=\frac{1}{\gamma_0^{k+1}}\sqrt{\frac{2}{\pi\chi^{\prime\prime}(\gamma_0)}}
\sqrt{\frac{1}{t_0}-\frac{1}{t}}\,m_k\,,
\label{eq:muk}
\ee
where the $m_k$'s are numerical constants. The first ones read
\be
\begin{split}
m_2&=
\frac{7\pi^2}{12}-\pi\ln 2+\ln^2 2
\\
m_3&=
\frac32\zeta(3)+
\frac{\pi^3}{8}-\frac{\pi^2}{4}\ln 2+\frac{3\pi}{2}\ln^2 2+\ln^3 2\\
m_4&=
3(2\ln 2-\pi)\zeta(3)
+\frac{119\pi^4}{240}
-\frac{\pi^3}{2}\ln 2
-\frac{\pi^2}{2}\ln^2 2
-2\pi\ln^3 2
+\ln^4 2
\end{split}
\ee
or in numbers,
$m_2=4.06013\cdots$,
$m_3=6.56570\cdots$,
$m_4=26.9902\cdots$.


\subsection{\label{sec:modelnum}Model suitable for a numerical implementation}

For simplicity of the implementation, we considered
a discretized branching diffusion
model. 
At each time step, a particle on lattice site $x$ 
(with lattice spacing $\Delta x=1$)
has the probability
$\Delta t$ to give birth to another particle on the same site, $\Delta t$ to move 
to the site $x+1$, $\Delta t$
to move to the site $x-1$, and $1-3\Delta t$ to stay unchanged at the
same site. 
The eigenfunctions of the corresponding diffusion kernel
are the exponential functions $e^{-\gamma x}$, and
the eigenvalues read
\be
\chi(\gamma)=\frac{1}{\Delta t}\ln
\left[
1+\Delta t\left(e^\gamma+e^{-\gamma}-1\right)
\right].
\ee
The discretization in time is chosen to be $\Delta t=0.01$.
The relevant parameters for this model are
\begin{equation}
\gamma_0=0.91338\cdots\ ,\ \ \chi^\prime(\gamma_0)=2.05412\cdots
\ ,\ \ \chi''(\gamma_0)=2.79893\cdots
\label{eq:numbers}
\end{equation}

\subsection{\label{sec:checkdet}Check of the deterministic analytical results}

We solve the equivalent of the deterministic 
FKPP equation with the critical initial condition. For our discretized model,
the FKPP equation becomes the finite difference equation
\be
l_{x+1}(t+\Delta t)=l_x(t)+\ln
\left\{1+\Delta t\left[e^{l_{x+1}(t)-l_x(t)}
+e^{l_{x-1}(t)-l_x(t)}-1-e^{l_x(t)}\right]\right\}
\ee
with the initial condition $l_x(t=0)=\ln\left[1-\exp(-e^{-\gamma_0 x})\right]$.
Here $x$ is an integer that labels the sites of the lattice.
$l_x(t)$ is the logarithm of the equivalent of $\phi$
defined in Sec.~\ref{sec:det}.
The use of a logarithmic variable avoids problems with
numerical accuracy in the region $\phi\rightarrow 0$, upon which the solution 
depends crucially.

\begin{figure}
\begin{center}
\includegraphics[width=.8\textwidth,angle=0]{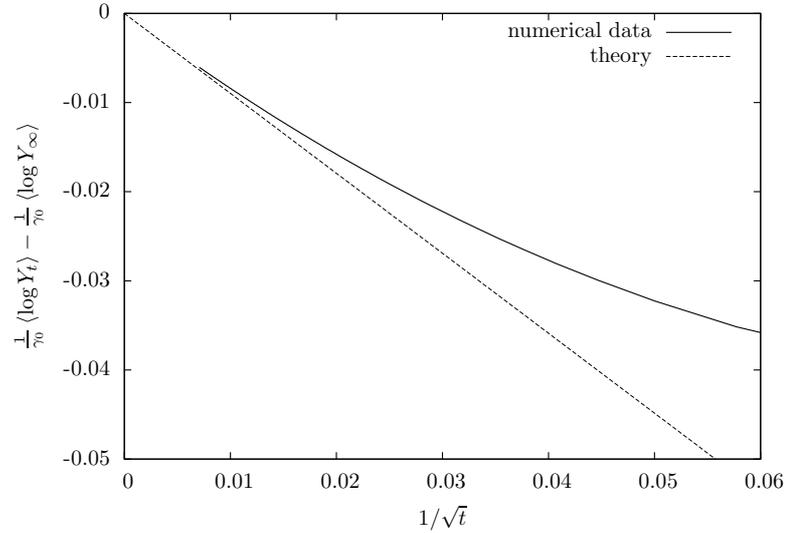}
\end{center}
\caption{\label{fig:plotpos}
$\frac{1}{\gamma_0}\langle \ln Y_t\rangle$
from the numerical solution of the FKPP equation with 
the ``critical'' initial condition,
as a function of $1/\sqrt{t}$.
(The constant term is subtracted.)
One sees that it converges to the analytical 
result Eq.~(\ref{eq:mu1fulldet}) (with $t_0\rightarrow +\infty$; straight line) 
for $t\rightarrow +\infty$.}
\end{figure}

First, 
we integrate the solution 
according to Eq.~(\ref{eq:logYfromphi})
in order to get $\frac{1}{\gamma_0}\langle\ln Y_t\rangle$ .
The analytical expectation for the model which is implemented is
given in Eq.~(\ref{eq:mu1fulldet}) with the numerical inputs~(\ref{eq:numbers}):
\be
\frac{1}{\gamma_0}\langle \ln Y_t\rangle=\text{const}-\frac{0.8969\cdots}{\sqrt
{t}}.
\ee
The numerical calculation is shown in Fig.~\ref{fig:plotpos}, 
and is in perfect agreement with the analytical formula.
In order to estimate more quantitatively the quality of this agreement,
we fit a function of the form 
\be
f(t)=c_0+\frac{c_{\frac12}}{\sqrt{t}}+
\frac{c_1}{t}+
\frac{c_{\frac32}}{t^{3/2}},
\label{eq:fit}
\ee 
where
the $c$'s are the free parameters.
The value of $c_{\frac12}$ which we get from the fit is $c_{\frac12}=0.8918$, which
is very close to the expected value from our analytical calculation.

\begin{figure}
\begin{center}
\includegraphics[width=.8\textwidth,angle=0]{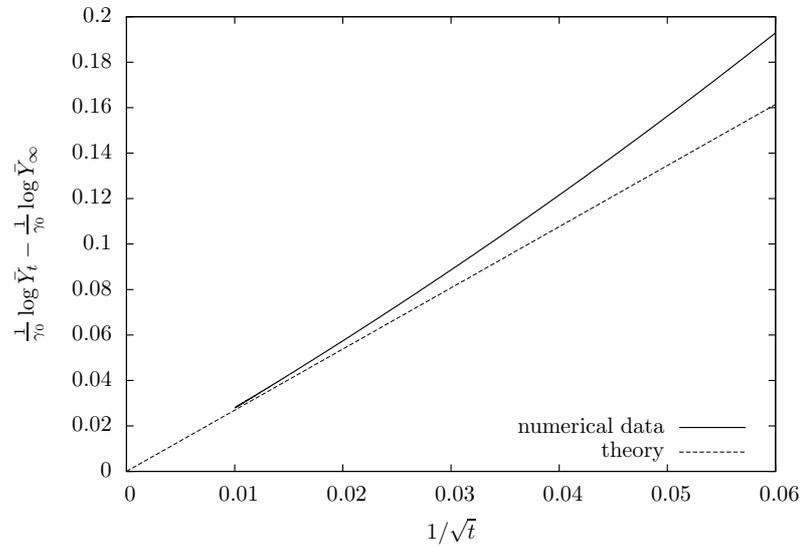}
\end{center}
\caption{\label{fig:plotcutoff}
$\frac{1}{\gamma_0}{\ln \bar Y_t}$
from the numerical solution of the branching diffusion equation with a cutoff
as a function of $1/\sqrt{t}$. (The constant is subtracted.)
Again, the numerical solution converges to the analytical result
(Eq.~(\ref{eq:logYdettip}); straight line) as $t$ gets large.
}
\end{figure}

Next, we solve the deterministic FKPP equation with a tip cutoff.
In practice, the latter cutoff is implemented as a smooth nonlinearity,
as in Eq.~(\ref{eq:tipcutoffsmooth}). More precisely,
the equation we solve numerically is
the following:
\be
l_x(t+\Delta t)=l_x(t)
+\ln\left\{1
+\Delta t\left[e^{l_{x+1}(t)-l_x(t)}+e^{l_{x-1}(t)-l_x(t)}-2
+\frac{1}{1+e^{-l_x(t)}}\right]\right\}
\ee
with $l_x(t=0)=-|x|$.
Here, $l_x(t)$ is the logarithm of the number of particles on site $x$ at time $t$.
The logarithmic scale for the evolved function is useful here because of
the exponential growth of the number of particles with time.
Also in this case, the result is in excellent agreement with the analytical expectation
(see Fig.~\ref{fig:plotcutoff}), which, for the
considered model, should read (see Eq.~(\ref{eq:logYdettip}))
\be
\frac{1}{\gamma_0}{ \ln \bar Y_t}=\text{const}
+\frac{2.6909\cdots}{\sqrt{t}}
\ee
The fit of the same function $f(t)$ as before
to the numerical data gives $c_{\frac12}=2.7120$ which, again,
is very close to the analytical estimate.

\subsection{\label{sec:checkstat}Check of the statistics of $f$}

We now use a Monte-Carlo implementation of the stochastic model of a
branching
random walk described above
in order to test the probability distribution of $f$ given 
in Eq.~(\ref{eq:pf}). 

The implementation is quite straightforward, except maybe 
that after a few timesteps,
the number of particles $n_x$ in the central bins 
(typically $|x|\leq \chi^\prime(\gamma_0) t$)
becomes very large.
To handle such large particle numbers, we further 
evolve these bins in a deterministic way.
(In practice in the code,
we set the limit between stochastic and deterministic evolution
at $n_x=10^6$.)
Such an approximate treatment was tested before in a similar context,
see e.g. Refs.~\cite{Moro:2004,BD:1999,BD:2001}.
As in the deterministic case discussed above, 
we also switch to logarithmic variables, $l_x\equiv\ln n_x$, in
order to be able to handle the large
particle numbers in a standard double-precision
computer representation.
Of course, the low-density tails of the system are treated fully stochastically.

\begin{figure}
\begin{center}
\includegraphics[width=.95\textwidth]{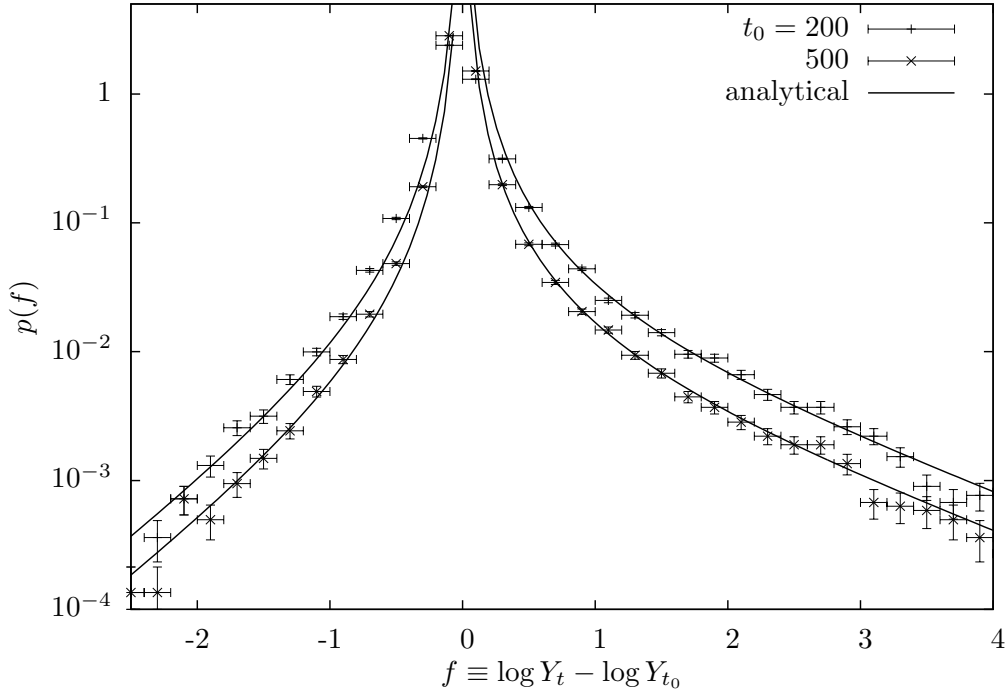}
\end{center}
\caption{\label{fig:plotdistributionf}
Distribution of $f$ for $t=1000$ and two different values of $t_0$. 
The numerical data (points with statistical error bars and bin width) 
are compared to Eq.~(\ref{eq:pf}) (continuous lines)
($\log_{10}$ scale on the vertical axis).
}
\end{figure}

The result for the distribution of $f$ is displayed in Fig.~\ref{fig:plotdistributionf}
compared to the analytical formulas~(\ref{eq:pf}).
We see an excellent agreement between the outcome of our model and the
numerical data.

\begin{figure}
\begin{center}
\includegraphics[width=.95\textwidth]{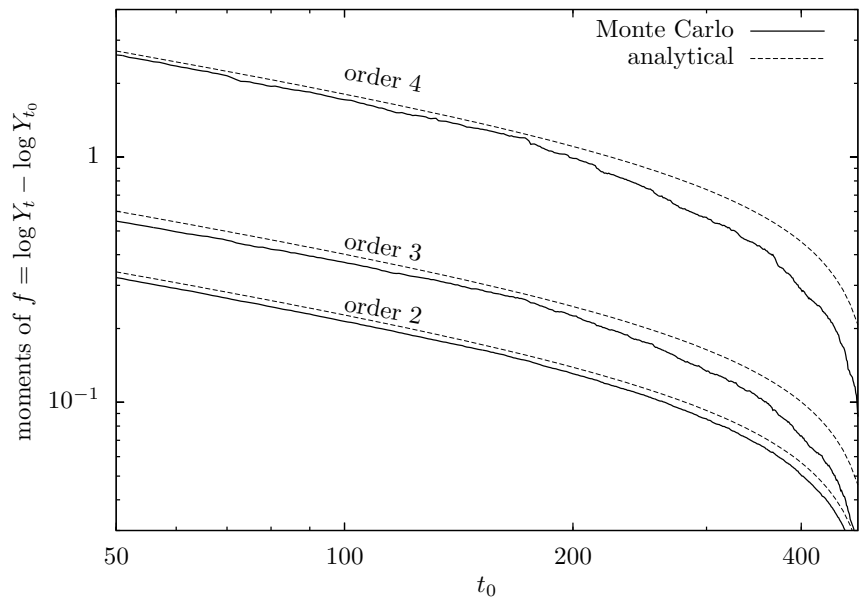}
\end{center}
\caption{\label{fig:plotmoments}
Moments of $f=\ln Y_t/Y_{t_0}$ of order 2 to 4 for $t=500$ as a function of $t_0$ in
logarithmic scales.
The numerical data (full lines) are compared to the analytical calculation
in Eq.~(\ref{eq:muk}) (dashed lines)
($\log_{10}$ scale on both axes).
There are about $4\times 10^5$  realizations in the statistical ensemble used
to perform the averages.
}
\end{figure}

We can also compute numerically the first few moments of the variable $f$
and plot them against $t_0$ (Fig.~\ref{fig:plotmoments}).
Here again, there is a good agreement between the analytical result
and the numerical calculation, although more statistics would be
needed in order to reach a good accuracy for the
moments of order 3 and 4.


\section{\label{sec:EVS}Stochastic interpretation of the $1/\sqrt{t}$ corrections
to the position of FKPP fronts}


So far, we have essentially discussed the statistics of the $\ln Y$
random variable in the light of our phenomenological picture of BRW.
We are now going to address the average of the position of the rightmost particle
$\langle X_t\rangle$ which is also the position of the FKPP front,
and whose expression at order $1/\sqrt{t}$ was obtained in Ref.~\cite{EVS:2000}.

\subsection{Correction to $\langle X_t\rangle$ due to fluctuations}

As in the case of $\langle f\rangle$, in our picture, the average value
of the position of the front is given by the deterministic evolution
of the bulk of the particles, supplemented by a contribution from fluctuations
in the
low-density region. We may write
\be
\mu_1^\prime\equiv\left\langle X_t\right\rangle-\left\langle X_{t_0}\right\rangle=
2(t-t_0)-\frac32\ln\frac{t}{t_0}
+C_{\bar X}
\left(\frac{1}{\sqrt{t}}-\frac{1}{\sqrt{t_0}}\right)
+\mu_1^{\prime +}-\mu_1^{\prime -}.
\label{eq:c1prime}
\ee
In this section, we shall compute $\mu_1^{\prime +}-\mu_1^{\prime -}$.
$\mu_1^{\prime +}$ is the contribution at time $t$ 
of fluctuations that occur all over the range of time
and $\mu_1^{\prime -}$ is the contribution at time $t_0$ of fluctuations that
have occurred before $t_0$:
\be
\mu_1^{\prime +}=\int_0^t dt_1\int_0^{+\infty} d\delta\,\delta X_t\ ,
\ \
\mu_1^{\prime -}=\int_0^{t_0} dt_1\int_0^{+\infty} d\delta\,\delta X_{t_0}\ ,
\ee
where the appropriate regulators will be introduced later.
$\delta X_t$ is the contribution to the shift of the position of the
tip of the front at time $t$ of
a fluctuation of size $\delta$ occurring at $t_1$.
Let us now evaluate $\delta X_t$.

When a fluctuation occurs at time $t_1$ at position $\delta$ ahead of the tip
$\bar X_{t_1}$ of the regular front, then it develops its own front
by independent branching diffusion.
The resulting density of particles at time $t\gg t_1$ becomes 
the sum of two terms,
and therefore has the shape
\be
\psi_{\bar X_t+\delta X_t}(x,t)=\psi_{\bar X_t}(x,t)+C\psi_{\bar X_{\delta,t}}(x,t)
\ee
where $\psi$ is given by Eq.~(\ref{eq:shape}) and $\bar X_{\delta,t}$ by Eq.~(\ref{eq:Xdelta}).
Using the latter equations, one is led to
\be
\delta X_t=\ln\left[1+C{e^\delta}\left(\frac{t}{t_1(t-t_1)}\right)^{3/2}\right].
\ee

The calculation of
$\mu_1^{\prime +}$ and $\mu_1^{\prime -}$
proceeds exactly as in the case of $\mu_1^+$ and $\mu_1^-$ 
in Sec.~\ref{sec:mu1}.
$\mu^{\prime+}_-$ is still given by an equation of the form of~(\ref{eq:mu1+}),
but with the replacements
${\cal I}_0\rightarrow {\cal I}^\prime_0$ and ${\cal I}_1\rightarrow {\cal I}^\prime_1$, 
where now
\be
{\cal I}_0^\prime=\int_{t_0^\prime/t}^{1-\bar t_0/t} d\lambda\lambda^{-3/2}(1-\lambda)^{-3/2}\ ,
\ \
{\cal I}_1^\prime=\int_{t_0^\prime/t}^{1-\bar t_0/t} d\lambda\lambda^{-3/2}(1-\lambda)^{-3/2}
\ln \left[\lambda^{3/2}(1-\lambda)^{3/2}\right].
\ee
Note that in the present case,
late times need to be cutoff in order to ensure the
convergence of the integrals:
We pick some arbitrary $\bar t_0\ll t$
say of order one.

The same change of variable as before may be used:
$\lambda=\sin^2\theta$, then
\be
{\cal I}_0^\prime=8\int_{\arcsin\sqrt\frac{t_0^\prime}{t}}^{\arcsin\sqrt{1-\frac{\bar t_0}{t}}}
\frac{d\theta}{\sin^2 2\theta}\ ,\ \
{\cal I}_1^\prime=24\int_{\arcsin\sqrt\frac{t_0^\prime}{t}}^{\arcsin\sqrt{1-\frac{\bar t_0}{t}}}
\frac{d\theta}{\sin^2 2\theta}\ln\frac{\sin 2\theta}{2}.
\ee
After performing the integrals and
expanding in the limit of small $\bar t_0$, $t_0^\prime$ compared to $t$, one gets
\be
\mu_1^{\prime+}=2CC_1\left\{
\frac{1}{\sqrt{\bar t_0}}\left(\ln \frac{\bar t_0^{3/2}}{C}+3\right)
+\frac{1}{\sqrt{t_0^\prime}}\left(\ln \frac{t_0^{\prime 3/2}}{C}+3\right)
-\frac{6\pi}{\sqrt{t}}
\right\}.
\ee
The main difference with respect to Eq.~(\ref{eq:c1+})
(once the relevant expansions have been performed)
is the presence of $\bar t_0$-dependent terms and of an extra factor 3
in the last term.
As before, $\mu_1^{\prime-}$ is deduced from the above formula
by replacing $t$ by $t_0$.
We then see that in the difference $\mu_1^{\prime+}-\mu_1^{\prime-}$, the
$\bar t_0$ and $t_0^\prime$ dependences cancel.

As for the moments of order $n\geq 2$,
they are found to depend on $\bar t_0$, that is,
on the late-time fluctuations, as they should,
since $X_t$ is the position of the rightmost particle, which
experiences a stochastic motion of size 1 over time scales of order 1.

\subsection{Recovering the
Ebert-Van Saarloos term}

Putting everything together, namely the value of $C_{\bar X}$ from Eq.~(\ref{eq:CMFX}) 
and the value of
$\mu_1^{\prime+}-\mu_1^{\prime-}$ just computed,
we find the interesting expression
\be
\left\langle X_{t}-X_{t_0}\right\rangle=
\mu_1^\prime=
2(t-t_0)-\frac32\ln\frac{t}{t_0}+
\left[
3\sqrt{\pi}
\left(
\frac{1}{\sqrt{t}}-\frac{1}{\sqrt{t_0}}
\right)\right]
-\left[6\sqrt{\pi}
\left(
\frac{1}{\sqrt{t}}-\frac{1}{\sqrt{t_0}}
\right)
\right].
\ee
The terms that grow with $t$ and $t_0$ are the usual deterministic terms
from Bramson's classical solution \cite{B:1983}.
Then, the next terms, under the square brackets, are respectively the
deterministic correction to the position of the discreteness cutoff in
the mean-field model, and the correction due to fluctuations.
We see that the latter is exactly twice the former, with a minus sign.
The sum of these two terms gives back the Ebert-Van Saarloos correction
for $\left\langle X_t-X_{t_0}\right\rangle$, see Eq.~(\ref{eq:FKPPpos}).

In other words, the mismatch between $\bar X_t$, the position of the tip
of the front in the deterministic model with a cutoff, and $\langle X_t\rangle$,
the mean position of the rightmost particle in the full stochastic model,
is exactly due to the very fluctuations we have been analyzing in this paper.


\section{\label{sec:conclusion}Conclusions}

Some time ago, we proposed a model for the fluctuations of stochastic pulled fronts
\cite{Brunet:2005bz}, which are realizations of the stochastic FKPP (sFKPP) equation 
(for a review, see Ref.~\cite{Panja:2004}).
Equations in the class of the sFKPP equation
may be thought of, for instance,
as representing the dynamics of the
particle number density in a branching-diffusion process
in which there is in addition a nonlinear selection/saturation process 
that effectively limits the density
of particles. The realizations of such
equations are stochastic traveling waves.
The stochasticity comes from the discreteness of the number of particles.
In this context, the (deterministic) FKPP equation represents the mean-field 
(or infinite number of particle) limit
of the full dynamics. 

Expansions about the mean-field solution
were considered already a long time ago; see, e.g., Ref.~\cite{BHP:1994}
where the so-called $\Omega$ expansion (see Ref.~\cite{VanKampen})
was applied to study fluctuations in the context of
reaction-diffusion processes.
Later, we could obtain new analytical results
thanks to a phenomenological
model~\cite{Brunet:2005bz}.
The picture encoded in our model was the following: 
Most of the time, the traveling wave 
front propagates deterministically,
obeying the ordinary deterministic
FKPP equation supplemented 
with a cutoff in the tail, accounting for discreteness by making sure that
the number density of particles reaches 0 rapidly whenever it drops below 1.
Brunet and Derrida had shown \cite{BD:1997} that such a cutoff
correctly represents the main effect of the noise on the velocity of the front.
On top of that, in our model,
there are some rare fluctuations consisting in a few particles
randomly sent far ahead of the tip of the front, which upon further evolution 
build up a new front that completely takes over the old one.
A positive correction to the front velocity was found, and the cumulants of the
front position were computed (see Ref.~\cite{Brunet:2005bz}).
\\

In the present paper, we have considered a simple branching random walk, without
any selection mechanism. We have used exactly the same ingredients as the ones
conjectured in the model for stochastic fronts, namely deterministic evolution with a cutoff
and fluctuations consisting in a few particles randomly sent ahead the 
tip of the front at a distance distributed exponentially.
We were also able to
arrive at a quantitative characterization of the fluctuations of the front
in these processes.
\\

There are however a few important differences between the branching random walk
and the stochastic FKPP front.
First, the initial fluctuations are never ``forgotten'' in the BRW case. 
This is because of the absence of a selection mechanism able to
``kill'' the front and let it be periodically regenerated by fluctuations.
Therefore,
we could only compute the effect of the fluctuations on the front position between two
large times $t_0$ and $t$.
Next, while it was quite straightforward 
to define a proper front position in the sFKPP case (as for example
the integral of the normalized particle density from position say 0 to $+\infty$), 
it is
more tricky for the simple branching random walk.
We were led to consider the variables $\ln Y$ and $\ln Z$ (introduced in
Sec.~\ref{sec:var}).
Our main result is the distribution of the variable $\ln Y_t/Y_{t_0}$ given
in Eq.~(\ref{eq:pf}), where
$t_0$ and $t$ are two large times such that $t_0,t,t-t_0\gg 1$.
Interestingly enough, the distribution of the positive values of this variable
is identical (up to an overall factor) to the distribution of the front
fluctuations in the sFKPP case.
The same holds true for the distribution of $\ln Z_t/Z_{t_0}$ to which we dedicate
Appendix~\ref{sec:Z}.

We were also able to discuss the average of the position of the rightmost particle, but
not its higher moments since they are sensitive to the very late-time fluctuations
which are not properly described in our model.
As for the average position, 
we could nevertheless propose an appealing interpretation of the ${\cal O}(1/\sqrt{t})$
correction to the front position computed by Ebert and Van Saarloos in Ref.~\cite{EVS:2000}.
\\

There are still many open questions.
Maybe the most outstanding one on the technical side would be to try and
compute the statistics of $\ln Y_t/Y_{t_0}$ (and of $\ln Z_t/Z_{t_0}$)
exactly,
instead of relying on a phenomenological picture involving conjectures. 
We outlined such a calculation in Appendix~\ref{sec:exact},
based on the evaluation of a generating function,
but without being able to complete it.

\section*{Acknowledgements}

The idea at the origin of Appendix~\ref{sec:exact} 
is due to Prof. B.~Derrida.
We thank him also for very helpful discussions, and for his reading
of the manuscript.
We acknowledge support from ``P2IO Excellence Laboratory'',
and from the US Department of Energy, Grant No. DE-FG02-92ER40699.

\appendix


\section{\label{sec:details}
Details of the calculation of the probability distribution of $f$}

In this appendix, we go back to the calculations that lead to 
Eqs.~(\ref{eq:shift}),~(\ref{eq:pf1}) and~(\ref{eq:pf2}),
but keeping the subleading terms that we neglected a priori 
in Sec.~\ref{sec:distrib1} in order to simplify the presentation.

The exact evaluation of $\bar Y_t$ starting from its definition given 
in Eq.~(\ref{eq:Y0}),
in which one inserts Eqs.~(\ref{eq:shape}),~(\ref{eq:Xt}),
makes use of the basic Gaussian integral
\be
\int_{-\infty}^0 dx\,(\alpha x+\beta)e^{-\frac{x^2}{4t}}=-2\alpha t+\beta\sqrt{\pi t}.
\ee
We immediately arrive at Eq.~(\ref{eq:Y0}),
which may also be rewritten at order $1/\sqrt{t}$ as
\be
\bar Y_t=-2\alpha e^{\frac{C_{\bar Y}}{\sqrt{t}}}.
\ee
We now add a fluctuation occurring say at time $t_1$.
It develops a front whose tip sits, at time $t$, at position
\be
\bar X_{\delta,t}=\bar X_t+\delta-\frac32\ln\frac{t_1(t-t_1)}{t}
+C_{\bar X}\left(
\frac{1}{\sqrt{t_1}}+\frac{1}{\sqrt{t-t_1}}-\frac{1}{\sqrt{t}}
\right),
\ee
which is Eq.~(\ref{eq:Xt}) supplemented with the subleading terms.
Keeping all the latter, we see that Eq.~(\ref{eq:shift}) just needs to
be replaced by
\be
\delta \ln Y_t=\ln
\left[
1+C \frac{e^\delta}{t_1^{3/2}}\sqrt{\frac{t}{t-t_1}}
e^{\frac{C_{\bar X}}{\sqrt{t_1}}+C_{\bar Y}\left(\frac{1}{\sqrt{t-t_1}}-\frac{1}{\sqrt{t}}\right)}
\right].
\ee
As for the probability distribution of the fluctuations in Eq.~(\ref{eq:pf01}), 
it becomes
\be
p(\delta f;t_1)=\frac{CC_1}{t_1^{3/2}}
\sqrt{\frac{t}{t-t_1}}
e^{\frac{C_{\bar X}}{\sqrt{t_1}}+C_{\bar Y}\left(\frac{1}{\sqrt{t-t_1}}-\frac{1}{\sqrt{t}}\right)}
\frac{e^{-\delta f}}{\left(1-e^{-\delta f}\right)^2},
\ee
which has to be integrated over $t_1$.
We recall that after integration over $t_1$,
the obtained expression will be correct at order $1/t_1$, $1/(t-t_1)$, $1/t$,
hence only the first nontrivial terms are relevant in the expansion of the
exponential.

In the absence of ${\cal O}(1/(t-t_1))$ terms, the integration
region could be chosen to be $[t_0,t]$ as in Sec.~\ref{sec:distrib}.
Now however we have a non-integrable singularity at $t_1=t$ which needs
to be cut off. Hence we write
\be
p(\delta f)=\int_{t_0}^{t-{\bar t_0}}dt_1\, p(\delta f;t_1)
=CC_1 e^{-\frac{C_{\bar Y}}{\sqrt{t}}}
\frac{e^{-\delta f}}{\left(1-e^{-\delta f}\right)^2}
\int_{t_0}^{t-{\bar t_0}}\frac{dt_1}{t_1^{3/2}}\sqrt{\frac{t}{t-t_1}}
e^{\frac{C_{\bar X}}{\sqrt{t_1}}+
\frac{C_{\bar Y}}{\sqrt{t-t_1}}}
\ee
where ${\bar t_0}$ is an arbitrary time interval 
whose length is of the order of~1.

Let us compute the integral 
\be
{\cal J}\equiv \int_{t_0}^{t-{\bar t_0}}\frac{dt_1}{t_1^{3/2}}\sqrt{\frac{t}{t-t_1}}
e^{\frac{C_{\bar X}}{\sqrt{t_1}}+
\frac{C_{\bar Y}}{\sqrt{t-t_1}}}
\ee
appearing in the previous expression. 
We expand the exponential to lowest order,
and hence we get the three terms
\be
{\cal J}= {\cal J}_0+C_{\bar X}{\cal J}_{1}^{(1)}+C_{\bar Y}{\cal J}_{1}^{(2)},
\ee
where ${\cal J}_0$
(which is essentially the same integral as
${\cal I}_0$ in Eq.~(\ref{eq:defI0}))
gives back the lowest-order result in Eq.~(\ref{eq:pf1}):
\be
{\cal J}_0=\int_{t_0}^{t-{\bar t_0}}
\frac{dt_1}{t_1^{3/2}}\sqrt{\frac{t}{t-t_1}}
=2\sqrt{\frac{t-t_0}{tt_0}}-2\sqrt{\frac{{\bar t_0}}{t(t-{\bar t_0})}}
\simeq 2\sqrt{\frac{t-t_0}{tt_0}}+{\cal O}(1/t).
\label{eq:I0exp}
\ee
As for the two other terms,
\be
{\cal J}_{1}^{(1)}= \sqrt{t}\int_{t_0}^{t-{\bar t_0}}\frac{dt_1}{t_1^{2}}\frac{1}{\sqrt{t-t_1}},\
{\cal J}_{1}^{(2)}= \sqrt{t}\int_{t_0}^{t-{\bar t_0}}\frac{dt_1}{t_1^{3/2}}{\frac{1}{t-t_1}}
\ee
are new contributions which are subleading, as is easy to demonstrate
from an exact calculation of these integrals.
We start with the computation of ${\cal J}_{1}^{(1)}$:
\be
{\cal J}_{1}^{(1)}=
\sqrt{\frac{t-t_0}{t}}
\left(
\frac{1}{t_0}
+\frac{1}{t}\arctanh\sqrt{\frac{t-t_0}{t}}
\right)
-\sqrt{\frac{{\bar t_0}}{t}}
\left(
\frac{1}{t-{\bar t_0}}
+\frac{1}{t}\arctanh\sqrt{\frac{{\bar t_0}}{t}}
\right).
\ee
Since $\arctanh\sqrt{1-x}\underset{x\rightarrow 0}{\sim}-\frac12\ln x$,
it is clear that the largest terms in ${\cal J}_{1}^{(1)}$ are
at most of order $\ln (t/t_0)/t$ and $1/t_0$.
As for ${\cal J}_{1}^{(2)}$,
\be
{\cal J}_{1}^{(2)}=
2\left(
\frac{1}{\sqrt{t t_0}}
-\frac{1}{\sqrt{t(t-{\bar t_0})}}
\right)
+\frac{2}{t}
\left(
\arctanh\sqrt{1-\frac{{\bar t_0}}{t}}
-\arctanh\sqrt{\frac{t_0}{t}}
\right)
\ee
The second term is divergent for ${\bar t_0}\rightarrow 0$. It gives the dominant contribution
at large $t$: ${\cal J}_{1}^{(2)}\sim \ln (t/{\bar t_0})/t$.
The other terms are also subleading, of order $1/\sqrt{t t_0}$ and $1/t$.

Hence we see that at order ${\cal O}(1/\sqrt{t},1/\sqrt{t_0},1/\sqrt{t-t_0})$, ${\cal J}$
boils down to the first term in the expansion 
of ${\cal J}_0$ in Eq.~(\ref{eq:I0exp}).

Lastly, we have already noticed in Sec.~\ref{sec:distrib} that
\be
f-\delta f=\ln \frac{\bar Y_t}{\bar Y_{t_0}}\simeq C_{\bar Y}
\left(\frac{1}{\sqrt{t}}-\frac{1}{\sqrt{t_0}}\right),
\ee
thus replacing $\delta f$ by $f$ in $p(\delta f)$ 
brings about only
subleading contributions.

All in all, we have justified the approximations that led to Eq.~(\ref{eq:pf1}).
From a very similar calculation, we would also recover Eq.~(\ref{eq:pf2}).


\section{\label{sec:Z}Statistics of $f_Z\equiv\ln Z_t-\ln Z_{t_0}$}

In the same way as for the variable $f=\ln Y_t-\ln Y_{t_0}$, we may try to
get the statistics of $f_Z=\ln Z_t-\ln Z_{t_0}$ from our phenomenological model.
The variable $Z_t$ is of interest since it is used in a mathematical theorem
to characterize what we call the position of the front in each realization,
however, as we shall see,
we cannot obtain full analytical formulas for the first moment of
$f_Z$ as in the case of $f$.
Moreover, as was already commented above, the $Z_t$ variable has properties
that make it awkward for numerical simulations.

The first step is to compute $Z_t$ in the mean-field approximation with a tip cutoff.
The result reads
\be
\bar Z_t=-2\alpha\sqrt{\pi}\left(1+\frac{3 \ln t}{2\sqrt{\pi t}}\right).
\ee
This formula is analogous to Eq.~(\ref{eq:Y0}), but there is now a
slightly stronger
$t$-dependence, $\propto\ln t/\sqrt{t}$, which we are able to determine
completely from the leading-order shape of the particle distribution.
We have dropped terms of order $1/\sqrt{t}$ and higher.

The effect of a fluctuation occurring at time $t_1$ on $\ln Z_t$ is
\be
\delta\ln Z_t=\ln\left[
1+C\frac{e^\delta}{t_1^{3/2}}
\frac{
1+\frac{\ln \left\{{[t_1(t-t_1)]^{\frac32}}{e^{-\delta}}\right\}}{\sqrt{\pi(t-t_1)}}}
{1+\frac{\ln t^{\frac32}}{\sqrt{\pi t}}}
\right]
\label{eq:shiftZ}
\ee
(Compare to Eq.~(\ref{eq:shift})).
The following approximate formula can now be written for the distribution of $f_Z$:
\be
p(f_Z)=\int dt_1\int_0^{+\infty}d\delta\,p(\delta)\,\delta\left[
f_Z-\left(\delta\ln Z_t-\delta\ln Z_{t_0}\right)
\right],
\ee
where $\delta \ln Z$ is given by Eq.~(\ref{eq:shiftZ}), while
$p(\delta)$ is the probability distribution~(\ref{eq:proba}).
The bounds on the integral over
$t_1$ depend on whether $f_Z$ is positive or negative.
Indeed, positive values of $f_Z$ are generated by fluctuations occurring
at $t_1$ between $t_0$ and $t$, while fluctuations before $t_0$
(namely between the times $t_0^\prime$ at which we declare that the system
contains a large number of particles and $t_0$) give rise to negative values
of $f_Z$.

The distribution of positive $f_Z$ is quite easy to compute.
It is enough to recognize that the terms of order $1/\sqrt{t-t_1}$ and $1/\sqrt{t}$
inside the
square bracket give subleading contributions to $p(f_Z)$.
It turns out that the final result is very similar to
$p(f)$ (see Eq.~(\ref{eq:pf1})), except for the detailed form of the $t_0$ and $t$ dependence:
\be
p(f_Z>0)=2CC_1
\left(
\frac{1}{\sqrt{t_0}}-\frac{1}{\sqrt{t}}
\right)
\frac{e^{-f_Z}}{\left(1-e^{-f_Z}\right)^2}.
\ee
Inserting the value of the constant $CC_1$ previously determined
(see Eq.~(\ref{eq:CC1}))
and going to a general branching diffusion kernel, we get
\be
p(f_Z>0)=
\sqrt{\frac{2}{\pi\chi^{\prime\prime}(\gamma_0)}}
\left(
\frac{1}{\sqrt{t_0}}-\frac{1}{\sqrt{t}}
\right)
\frac{e^{-f_Z}}{\left(1-e^{-f_Z}\right)^2}.
\label{eq:pfZ}
\ee

Negative values of $f_Z$ are more complicated to deal with
since we can no longer neglect the  $1/\sqrt{t-t_1}$
term in Eq.~(\ref{eq:shiftZ}) a priori.
Performing the change of variable $u\equiv t_1^{3/2}/(C e^\delta)$
and expanding for large $t$ and $t_0$, the equation for $p(f_Z)$ simplifies to
\be
p(f_Z<0)=CC_1\int_{t_0^\prime}^{t_0}\frac{dt_1}{t_1^{3/2}}
\int_0^{\frac{t_1^{3/2}}{C}}
du\,
\delta\left[
f_Z-\frac{1}{\sqrt{\pi}}\frac{\ln \left(Cu\right)}{1+u}\left(
\frac{1}{\sqrt{t}}-\frac{1}{\sqrt{t_0}}
\right)
\right].
\ee
Due to the Dirac $\delta$-function,
we see that
$p(f_Z)=0$ as soon as
$f_Z<\frac{1}{\sqrt{\pi} u_0}\left(\frac{1}{\sqrt{t}}-\frac{1}{\sqrt{t_0}}\right)$,
where $u_0$ solves
$\ln\left(Cu_0\right)=1+\frac{1}{u_0}$, and hence
is of order~1.
This means that $p(f_Z)$ is of higher-order in powers of $1/\sqrt{t}$
and $1/\sqrt{t_0}$ when $f_Z<0$.

\begin{figure}
\begin{center}
\includegraphics[width=.95\textwidth]{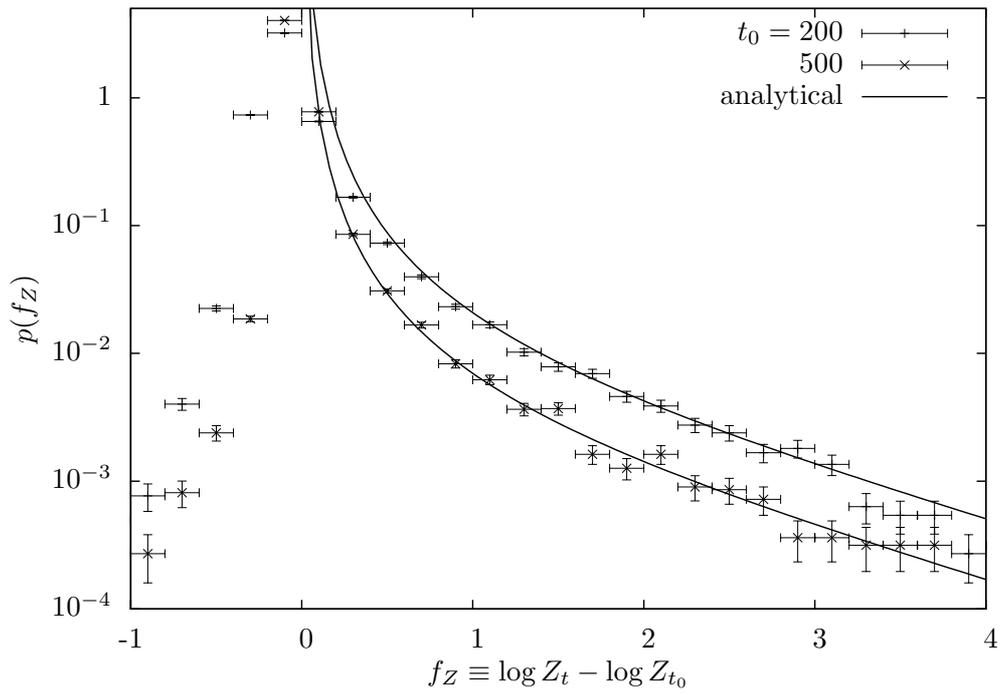}
\end{center}
\caption{\label{fig:plotdistributionfZ}
Distribution of $f_Z$ for $t=1000$ and two different values of $t_0$. 
The numerical data (points with statistical error bars and bin width) 
are compared to Eq.~(\ref{eq:pfZ}) (continuous lines).
($\log_{10}$ scale on the vertical axis).
}
\end{figure}

Our formula for the distribution, Eq.~(\ref{eq:pfZ}), successfully
compares to the numerical data, see Fig.~\ref{fig:plotdistributionfZ}.
We also see that the distribution of negative values of $f_Z$ is indeed
sharply suppressed (compare to the distribution of $f$ in
Fig.~\ref{fig:plotdistributionf}).\\

As for the mean of $f_Z$, we found that it depends on the arbitrary time
$t_0^\prime$ roughly as $1/\sqrt{t_0^\prime}$, and thus is not calculable.


\section{\label{sec:exact}Generating function for the moments}

In this section, we are going to find the form of the large positive 
$f$-fluctuations
from a generating function, hence from a deterministic calculation.

\subsection{General framework and exact formulas}

We can write the following identity:
\be
\left\langle
\left(\frac{\tilde Y_t}{\tilde Y_{t_0}}\right)^\nu
\right\rangle
=\frac{\sin\pi\nu}{\pi\nu}
\int_0^{+\infty}
{d\tilde u}
{d\tilde v}
\left(\frac{\tilde v}{\tilde u}\right)^\nu
\frac{\partial^2}{\partial \tilde u\partial \tilde v}
\left\langle
e^{-\tilde u \tilde Y_t-\tilde v \tilde Y_{t_0}}
\right\rangle,
\label{eq:momentsexacts}
\ee
see Eq.~(\ref{eq:defYt}) for the definition of $\tilde Y$.
This equation follows from the integral representation of the $\Gamma$
function, and
is suitable for series expansions
in $\nu$, which eventually lead to the moments of $\ln \tilde Y_t/\tilde Y_{t_0}$.
For some calculations outlined below, it will prove useful
to change $\tilde u$ and $\tilde v$ to the variables
\be
u=\tilde u \frac{e^{2t}}{\sqrt{t}}\ ,
\ \
v=\tilde v \frac{e^{2t_0}}{\sqrt{t_0}}
\ee
since Eq.~(\ref{eq:momentsexacts}) then holds in the very same form
(just up to the replacements
$\{\tilde u,\tilde v,\tilde Y_t,\tilde Y_{t_0}\}\rightarrow
\{u,v,Y_t,Y_{t_0}\}$)
directly for the moments of $\ln Y_t/Y_{t_0}=f$.

Let us introduce the generating function
\be
G_{t_0}(x)=\left\langle
e^{-\left(\tilde u \tilde Y_t+\tilde v \tilde Y_{t_0}\right)e^{-x}}
\right\rangle.
\ee
It is the value of this function at zero, 
$G_{t_0}(0)$, from which one computes the generating function
in Eq.~(\ref{eq:momentsexacts}), which reads
\be
\left\langle
\left(\frac{\tilde Y_t}{\tilde Y_{t_0}}\right)^\nu
\right\rangle
=\frac{\sin\pi\nu}{\pi\nu}
\int_0^{+\infty}
{d\tilde u}
{d\tilde v}
\left(\frac{\tilde v}{\tilde u}\right)^\nu
\frac{\partial^2G_{t_0}(0)}{\partial \tilde u\partial \tilde v}.
\label{eq:momentsexacts2}
\ee
The function $G_{t_0}(x)$ may also be written as
\be
G_{t_0}(x)=\left\langle
\prod_{i=1}^{N(t_0)} g_{\tau}(x-x_i(t_0))e^{-\tilde v e^{-(x-x_i(t_0))}}
\right\rangle,
\label{eq:Gexact}
\ee
where $\tau\equiv t-t_0$ is a parameter in this equation, and
\be
g_\tau(x)\equiv \left\langle e^{-\tilde u\tilde Y_\tau e^{-x}}\right\rangle.
\ee
In this form, it is clear that $G_{t_0}(x)$ obeys the FKPP equation (with 
time variable $t_0$), with the initial condition $g_\tau(x)e^{-\tilde v e^{-x}}$.
But $g_\tau(x)$ may also be written as
\be
g_\tau(x)=\left\langle \prod_{i=1}^{N(\tau)}e^{-\tilde u e^{-(x-x_i(\tau))}}\right\rangle,
\ee
which makes it obvious that it
also obeys the FKPP equation (with time variable $\tau$), 
with the initial
condition $g_0(x)=e^{-\tilde u e^{-x}}$.

So far, these formulas are exact and should in principle enable
the computation of the moments of $f$, from some hopefully limited
 knowledge of the
properties of the solutions to the FKPP equation.

We have not been able to fully compute the generating function. 
However, we
can use the systematic solution to FKPP for the evolution of $g$, and
a mean-field approximation for $G$: Interestingly enough, this turns out to be
enough to compute the positive fluctuations of $f$.

\subsection{Approximate solution: Moments of $f>0$}

In this section, we shall consider the stronger limit $t\gg t_0\gg 1$.

Let us treat the evolution from the initial time $t=0$ to time $t_0$
in the mean-field approximation with a tip cutoff:
This means that we assume a distribution of particles at time $t_0$ given
by Eq.~(\ref{eq:shape}).
Then the product over the particles in Eq.~(\ref{eq:Gexact}) becomes the 
exponential of an integral over the spatial coordinate weighted by
the particle density:
\be
G_{t_0}(x)=\exp
\left[
-\int_{-\infty}^{\bar X_{t_0}}dx^\prime
\alpha (x^\prime-\bar X_{t_0})
e^{-(x^\prime-\bar X_{t_0})-\frac{(x^\prime-\bar X_{t_0})^2}{4t_0}}
\left\{
\tilde v\,e^{-(x-x^\prime)}-\ln \left[g_{t-t_0}(x-x^\prime)\right]
\right\}
\right],
\label{eq:Gapprox}
\ee
where $\bar X_{t_0}=2t_0-\frac32\ln t_0$.
We have dropped the $\beta$ term in the form of the particle
distribution as well as the $1/\sqrt{t_0}$ term in $\bar X_{t_0}$ 
since they would eventually give 
subleading contributions,
of order $1/t_0$, at large $t_0$.

We see that the Gaussian under the integral makes sure that
the range of integration in the variable $x^\prime-\bar X_{t_0}$
is effectively $[-2\sqrt{t_0},0]$.

We turn to the $g_{t-t_0}$.
We know that it obeys the FKPP equation with the critical initial condition.
Hence the solution can be deduced from Eq.~(\ref{eq:phi}).
However, since $t_0\ll t$, defining
$\xi=x-\ln u-2(t-t_0)+\frac12\ln(t-t_0)$,
we may expand the solution for $1\ll \xi\ll\sqrt{t-t_0}$, namely
\be
1-g_{t-t_0}(\xi)\simeq \frac{1}{\sqrt{\pi}}\,\xi\, e^{-\xi}.
\ee
We have dropped the term of order $1/\sqrt{t-t_0}$ in $\xi$.

We shall now proceed with the integration in Eq.~(\ref{eq:Gapprox}).
Keeping only the term of order $1/\sqrt{t_0}$ and switching to the $u$, $v$ variables, 
we find
\be
G_{t_0}(0)=e^{-2(u+v)}\left(
1+u\frac{2\ln u-3\ln t_0}{\sqrt{\pi t_0}}
\right).
\ee
Inserting this expression into Eq.~(\ref{eq:momentsexacts2}) 
(with $\tilde u$, $\tilde v$
being replaced by $u$, $v$),
we now perform the integrals over $u$ and $v$.
The exact result is
\be
\left\langle
e^{\nu f}
\right\rangle
=
1+\frac{1}{\sqrt{\pi t_0}}
\left[
-\nu\,\psi(-\nu)+1+\nu\left(\frac32\ln t_0+\ln 2\right)
\right].
\ee
Remarkably, if we
invert this equation for the probability 
distribution
of $f$ by performing an appropriate contour
integration over $\nu$, we exactly recover Eq.~(\ref{eq:pf}) for the case $f>0$
(in the limit $t\rightarrow+\infty$, and up to replacements of the parameters
in~(\ref{eq:pf}): $\gamma_0\rightarrow 1$, 
$\chi^{\prime\prime}(\gamma_0)\rightarrow 2$).
Note that the constant $CC_1$ which appeared in the phenomenological model
is determined without any further calculation in the present approach.
The case $f<0$ however cannot be obtained unless we were able to
release the mean-field approximation for 
the evolution between $t=0$ and $t=t_0$.


\end{document}